\documentstyle[epsfig]{mn}

\newif\ifAMStwofonts

\def \eg{e.g.,\/}
\def \ie{i.e.,\/}
\def \Hb{${\rmn H}\beta$}
\def \HgA{${\rmn H}\gamma_{\rmn A}$}
\def \Hg{${\rmn H}\gamma$}
\def \HgF{${\rmn H}\gamma_{\rmn F}$}
\def \mgb{Mg{\,\it b}}
\def \mgbp{Mg{\,\it b}\,$^\prime$}
\def \kms{\rm km~s$^{-1}$}
\def \kmsM{\rm km~s$^{-1}$~Mpc$^{-1}$}
\def \prim{$^\prime$}

\ifoldfss
  \newcommand{\rmn}[1] {{\rm #1}}

  \ifCUPmtlplainloaded \else
    \NewTextAlphabet{textbfit} {cmbxti10} {}
    \NewTextAlphabet{textbfss} {cmssbx10} {}
    \NewMathAlphabet{mathbfit} {cmbxti10} {} 
    \NewMathAlphabet{mathbfss} {cmssbx10} {} 
  \fi
  \ifAMStwofonts
    \ifCUPmtlplainloaded \else
      \NewSymbolFont{upmath} {eurm10}
      \NewSymbolFont{AMSa} {msam10}
      \NewMathSymbol{\upi}     {0}{upmath}{19}
      \NewMathSymbol{\umu}     {0}{upmath}{16}
      \NewMathSymbol{\upartial}{0}{upmath}{40}
      \NewMathSymbol{\leqslant}{3}{AMSa}{36}
      \NewMathSymbol{\geqslant}{3}{AMSa}{3E}

       \let\le=\leqslant
       \let\ge=\geqslant
    \fi
  \fi
\fi 

\ifnfssone
  \newmathalphabet{\mathit}
  \addtoversion{normal}{\mathit}{cmr}{m}{it}
  \addtoversion{bold}{\mathit}{cmr}{bx}{it}
  \newcommand{\rmn}[1] {\mathrm{#1}}

  \newmathalphabet{\mathbfit} 
  \addtoversion{normal}{\mathbfit}{cmr}{bx}{it}
  \addtoversion{bold}{\mathbfit}{cmr}{bx}{it}
  \newmathalphabet{\mathbfss} 
  \addtoversion{normal}{\mathbfss}{cmss}{bx}{n}
  \addtoversion{bold}{\mathbfss}{cmss}{bx}{n}
  \ifAMStwofonts
    \ifCUPmtlplainloaded \else
      \UseAMStwoboldmath
      \makeatletter
      \new@mathgroup\upmath@group
      \define@mathgroup\mv@normal\upmath@group{eur}{m}{n}
      \define@mathgroup\mv@bold\upmath@group{eur}{b}{n}
      \edef\UPM{\hexnumber\upmath@group}
      \new@mathgroup\amsa@group
      \define@mathgroup\mv@normal\amsa@group{msa}{m}{n}
      \define@mathgroup\mv@bold\amsa@group{msa}{m}{n}
      \edef\AMSa{\hexnumber\amsa@group}
      \makeatother
      \mathchardef\upi="0\UPM19
      \mathchardef\umu="0\UPM16
      \mathchardef\upartial="0\UPM40
      \mathchardef\leqslant="3\AMSa36
      \mathchardef\geqslant="3\AMSa3E

       \let\le=\leqslant
       \let\ge=\geqslant
    \fi
  \fi
\fi 

\ifnfsstwo
  \newcommand{\rmn}[1] {\mathrm{#1}}

  \DeclareMathAlphabet{\mathbfit}{OT1}{cmr}{bx}{it}
  \SetMathAlphabet\mathbfit{bold}{OT1}{cmr}{bx}{it}
  \DeclareMathAlphabet{\mathbfss}{OT1}{cmss}{bx}{n}
  \SetMathAlphabet\mathbfss{bold}{OT1}{cmss}{bx}{n}
  \ifAMStwofonts
    \ifCUPmtlplainloaded \else
      \DeclareSymbolFont{UPM}{U}{eur}{m}{n}
      \SetSymbolFont{UPM}{bold}{U}{eur}{b}{n}
      \DeclareSymbolFont{AMSa}{U}{msa}{m}{n}
      \DeclareMathSymbol{\upi}{0}{UPM}{"19}
      \DeclareMathSymbol{\umu}{0}{UPM}{"16}
      \DeclareMathSymbol{\upartial}{0}{UPM}{"40}
      \DeclareMathSymbol{\leqslant}{3}{AMSa}{"36}
      \DeclareMathSymbol{\geqslant}{3}{AMSa}{"3E}

       \let\le=\leqslant
       \let\ge=\geqslant
    \fi
  \fi
\fi 

\ifCUPmtlplainloaded \else
  \ifAMStwofonts \else 
    \def\upi{\pi}
    \def\umu{\mu}
    \def\upartial{\partial}
  \fi
\fi

\title{The stellar populations of early-type galaxies in the Fornax
  cluster}

\author[Harald Kuntschner]
{Harald Kuntschner\thanks{email: harald.kuntschner@durham.ac.uk}\\
  University of Durham, Department of Physics, South Road, Durham DH1
  3LE, UK} 

\date{Jan 10, 2000, accepted for publication in Mon.\ Not. Royal Astron.\ Soc.}

\pagerange{\pageref{firstpage}--\pageref{lastpage}}
\pubyear{2000}

\begin{document}

\maketitle

\label{firstpage}

\begin{abstract}
  We have measured central line strengths for a magnitude-limited
  sample of early-type galaxies in the Fornax cluster, comprising 11
  elliptical (E) and 11 lenticular (S0) galaxies, more luminous than
  $\rm{M}_B=-17$. When compared with single-burst stellar population
  models we find that the centres of Fornax ellipticals follow a locus
  of fixed age and have metallicities varying roughly from half solar
  to two times solar. The centres of (lower luminosity) lenticular
  galaxies, however, exhibit a substantial spread to younger
  luminosity-weighted ages indicating a more extended star formation
  history.
  
  Galaxies with old stellar populations show tight scaling relations
  between metal-line indices and the central velocity dispersion.
  Remarkably also the Fe-lines are well correlated with $\sigma_0$. Our
  detailed analysis of the stellar populations suggests that these
  scaling relations are driven mostly by metallicity. Galaxies with a
  young stellar component do generally deviate from the main relation.
  In particular the lower luminosity S0s show a large spread.
  
  Our conclusions are based on several age/metallicity diagnostic
  diagrams in the Lick/IDS system comprising established indices such
  as Mg$_2$ and H$\beta$ as well as new and more sensitive indices such
  as H$\gamma_{\rm{A}}$ and Fe3, a combination of three prominent
  Fe-lines. The inferred difference in the age distribution between
  lenticular and elliptical galaxies is a robust conclusion as the
  models generate consistent relative ages using different age and
  metallicity indicators, even though the absolute ages remain
  uncertain. The absolute age uncertainty is mainly caused by the
  effects of non-solar abundance ratios which are not yet accounted for
  by the stellar population models. 
  
  Furthermore we find that elliptical galaxies and the bulge of one
  bright S0 are overabundant in magnesium, where the most luminous
  galaxies show the strongest overabundances. The stellar populations
  of young and faint S0s are consistent with solar abundance ratios or
  a weak Mg underabundance. Two of the faintest lenticular galaxies in
  our sample have blue continua and extremely strong Balmer-line
  absorption suggesting star formation $< 2$~Gyrs ago.

\end{abstract}

\begin{keywords}
  galaxies: abundances - galaxies: clusters: individual: Fornax -
  galaxies: formation - galaxies: elliptical and lenticular - galaxies:
  kinematics and dynamics
\end{keywords}

\section{INTRODUCTION}
\label{sec:intro}
Great efforts have been made in the last few years to develop
evolutionary stellar population synthesis models
\cite{bru93,wor94a,wei95,vaz96,kod97} in order to analyze the
integrated light of galaxies and derive estimates of their mean ages
and metal abundances. One of the main obstacles in the interpretation
has been the age/metallicity degeneracy in old stellar populations. As
pointed out by Worthey~(1994) the integrated spectral energy
distribution (SED) of an old ($>2$~Gyrs) stellar population looks
almost identical when the age is doubled and total metallicity reduced
by a factor of three at the same time.  Therefore two galaxies with
almost identical broad-band colours can have significantly different
ages and metallicities. In the optical wavelength range, only a few
narrow band absorption line-strength indices and the 4000~\AA\/ break
(see also Gorgas et~al. 1999) have so far been identified which can
break this degeneracy.

One of the most successful and widely used methods for measuring the
strength of age/metallicity discriminating absorption features is the
Lick/IDS system \cite{bur84,wor94b,tra98} which has been used by many
authors \cite{gon93,dav93,fis95,fis96,zie97,lon98,meh98,jor99}. In
contrast with high resolution index systems \cite{ros94,jon95}, which
promise a better separation of age and metalllitiy, the Lick/IDS system
allows the investigation of dynamically hot galaxies that have
intrinsically broad absorption lines. By plotting an age sensitive
index and a metallicity sensitive index against each other, one can
(partially) break the age/metallicity degeneracy and estimate, with the
help of model predictions, the luminosity weighted age and metallicity
of an integrated stellar population (see Figure~\ref{fig:hb_metal}).
Most recently, J{\o}rgensen (1999) used this methodology to investigate
the stellar populations of a large sample of early-type galaxies in the
Coma cluster. She concluded that there are real variations in both the
ages and the abundances while an anti-correlation between the mean ages
and the mean abundances makes it possible to maintain a low scatter in
scaling relations such as Mg--$\sigma_0$. Colless et~al. (1999) present
similar conclusions from the analysis of a combination of the
Mg--$\sigma_0$ relation and the Fundamental Plane in a large sample of
cluster early-type galaxies.

The spread in the ages for early-type galaxies and the anti-correlation
of age and metallicity found by the previous authors supports the
hierarchical picture for the construction of galaxies in which galaxies
form via several mergers involving star-formation \cite{bau96,kau96}.
However, the results are inconsistent with the conventional view that
all luminous elliptical galaxies are old and coeval. In this picture
the global spectrophotometric relations observed for ellipticals, for
example the colour-magnitude relation \cite{vis77,bow92,ter98} are
explained by the steady increase in the abundance of heavy elements
with increasing galaxy mass. This increase arises naturally in galactic
wind models such as that of Arimoto \& Yoshii (1987) and Kodama \&
Arimoto (1997).

Although with line-strength indices we can (partially) break the
age/metallicity degeneracy this is by no means the last obstacle to
overcome on our way to fully understand the stellar populations of
early-type galaxies and the cause of scaling ralations. Since the late
70's evidence has been accumulating that abundance ratios in galaxies
are often non-solar. In particular the Magnesium to Iron ratio seems to
be larger in luminous early-type galaxies when compared to solar
neighbourhood stars \cite{oco76,pel89,wor92,dav93,hen99,jor99}.
However, with only a very few exceptions (\eg\/ Weiss, Peletier \&
Matteucci 1995), non-solar abundance ratios have not yet been
incorporated in the model predictions. Among other issues this seems to
be the most important single problem which prevents us from deriving
accurate {\em absolute} age and metallicity estimates from integrated
light spectroscopy \cite{wor98}. Nevertheless with the current models
and high S/N data we are able to study relative trends in ages and
abundances as well as start to investigate the effects of non-solar
abundance ratios for individual elements \cite{wor98,pel99}.

In this paper, high S/N nuclear spectra of a complete sample of
early-type galaxies in the Fornax cluster brighter than $M_B=-17$ are
analyzed in the Lick/IDS system. The early results of this study have
already been presented in a letter to this journal \cite{kun98a}. This
paper is organized as follows. Section~\ref{sec:data} describes the
sample selection and basic data reduction. The calibration of the
line-strength indices to the Lick/IDS system is presented in
Section~\ref{sec:lick_cal} and the Appendix.
Section~\ref{sec:consistent} presents a consistency test for the model
predictions and our measured line-strength indices. In
Section~\ref{sec:nucstell} several index combinations are compared to
model predictions. In particular the effects of non-solar abundance
ratios, composite stellar populations and age/metallicity estimates of
the integrated light are discussed. In Section~\ref{sec:linesigma}
observed index--$\sigma_0$ relations are presented. Relations between
derived parameters such as age, metallicity and [Mg/Fe] with the
central velocity dispersion are investigated in
Section~\ref{sec:derived_sigma}. We then discuss the implications of
our results in Section~\ref{sec:discuss} and present the conclusions in
Section~\ref{sec:conclusion}. The fully corrected Lick/IDS indices of
our sample are tabulated in the Appendix.

\section{THE OBSERVATIONS AND DATA REDUCTION} 
\label{sec:data}
\subsection{The sample}
Our sample of 22 early-type galaxies has been selected from the
catalogue of Fornax galaxies (Ferguson 1989, hereafter F89), in order
to obtain a complete sample down to $B_T = 14.2$ or
$M_B=-17.$\footnote{Adopting a distance modulus of $m-M=31.2$; based on
  I-band surface brightness fluctuations \cite{jen98}. This corresponds
  to H$_0 \simeq 80$ \kmsM\/ for a flat Universe.} We have adopted the
morphological classifications given by F89 and checked them with images
we obtained on the Siding Spring 40$\arcsec$ telescope. From these we
noted a central dust lane in ESO359-G02 and a central disc in
ESO358-G59 which led us to classify them as lenticular galaxies. NGC
1428 was not observed because of a bright star close to its centre. We
also added the elliptical galaxy IC2006 to our sample, as it was not
classified by F89. The bona-fide elliptical NGC3379 was observed as a
calibration galaxy. The observations were carried out with the AAT
(3.9m) on the nights of 1996 December 6-8 using the RGO spectrograph.
The characteristics of the detector and the instrument set-up is given
in Table~\ref{tab:instr96}.

\begin{table}
  \caption[]{The instrumental set-up}
  \label{tab:instr96} 
  \begin{tabular}{ll} \hline 
    Telescope            &AAT (3.9m)                           \\
    Dates                &6-8 December 1996                    \\ 
    Instrument           &RGO spectrograph                     \\ \hline 
    Spectral range       &4243 - 5828 \AA                      \\
    Grating              &600 V                                \\
    Dispersion           &1.55 \AA pixel$^{-1}$                \\
    Resolution (FWHM)    &$\sim 4.1$ \AA                       \\   
    Spatial Scale        &0\farcs77 pixel$^{-1}$               \\
    Slit Width           &2\farcs3                              \\
    Detector             &Tek1k \#2 (24 $\umu$m$^2$ pixels)    \\
    Gain                 &$1.36\ e^- \mbox{ADU}^{-1}$        \\
    Read-out-noise       &$3.6\ e^-$ rms                       \\ 
    Seeing               &$\sim$1\arcsec                        \\ \hline
  \end{tabular} 
\end{table}

Typically, exposure times were between 300 and 1800 sec per galaxy
(see Table~\ref{tab:log_gal} for a detailed listing). For most of the
observations the slit was centred on the nucleus at ${\rmn PA} =
90^\circ$. The seeing was generally better than one arcsec.
Additionally we observed 15 different standard stars (mainly K-giants)
during twilight to act as templates for velocity dispersion
measurements as well as to calibrate our line-strength indices to the
Lick/IDS system \cite{wor94b}. The spectrophotometric standard stars
GD~108 and L745-46A were observed to enable us to correct the
continuum shape of our spectra. Table~\ref{tab:log_stars} lists all
observed standard stars with their spectral types (obtained from {\tt
SIMBAD}, operated by {\tt CDS}, Strasbourg) and also comments on their
use as Lick/IDS standard, velocity standard or spectrophotometric
standard.

\begin{table}
  \caption[]{Log of observations: galaxies}
  \label{tab:log_gal} 
  \begin{tabular}{llrrr}\hline 
Galaxy      &Type   &B$_T$ &Exp. time&PA       \\
            &       &[mag] &[sec]      &[\degr]\\\hline 
NGC 1316    &S0 pec &  9.4 & 300     & 90      \\  
NGC 1336    &E4     & 13.1 &1200     & 90      \\
NGC 1339    &E4     & 12.5 & 600     & 90      \\
NGC 1351    &E5     & 12.5 & 900     & 90      \\ 
NGC 1373    &E3     & 14.1 &1200     & 90      \\
NGC 1374    &E0     & 12.0 & 900     & 90      \\  
NGC 1375    &S0     & 13.2 & 900     & 90      \\  
NGC 1379    &E0     & 11.8 & 600     & 90      \\  
            &       &      & 600     & 90      \\  
NGC 1380    &S0     & 10.9 & 300     & 90      \\  
NGC 1380A   &S0     & 13.3 & 900     & 90      \\  
NGC 1381    &S0     & 12.4 & 480     & 90      \\  
            &       &      & 900     &140      \\  
            &       &      & 900     & 50      \\  
NGC 1381$^a$&S0     &      &1800     & 50      \\  
NGC 1399    &E0, cD & 10.6 & 300     & 90      \\
            &       &      & 600     &181      \\
NGC 1404    &E2     & 11.0 & 300     & 90      \\
            &       &      & 300     & 90      \\
NGC 1419    &E0     & 13.5 &1200     & 90      \\  
NGC 1427    &E4     & 11.8 & 600     & 90      \\
            &       &      &1800     & 79      \\
            &       &      &1800     &169      \\
IC 1963     &S0     & 12.9 & 600     & 90      \\  
IC 2006     &E      & 12.2 & 600     & 90      \\  
ESO 359-G02 &S0     & 14.2 &1200     & 90      \\ 
ESO 358-G06 &S0     & 13.9 &1800     & 90      \\  
ESO 358-G25 &S0 pec & 13.8 &1200     & 90      \\  
ESO 358-G50 &S0     & 13.9 &1200     & 90      \\  
ESO 358-G59 &S0     & 14.0 &1200     & 90      \\ \hline
NGC 3379$^b$&E1     & 10.2 & 300     & 90      \\ \hline
  \end{tabular} 

\medskip
Notes -- $^a$offset from nucleus; $^b$calibration
galaxy, non Fornax member
\end{table}

\begin{table}
  \caption[]{Log of observations: stars}
  \label{tab:log_stars}
  \begin{tabular}{lll} \hline 
Name     &Type           &comment  \\ \hline 
HD004656 &K4IIIb         &Lick/IDS std \\ 
HD037160 &K0IIIb         &Lick/IDS std \\ 
HD040657 &K1.5III        &velocity std \\ 
HD047205 &K1III          &Lick/IDS std \\ 
HD050778 &K4III          &Lick/IDS std \\ 
HD054810 &K0III          &Lick/IDS std \\ 
HD058972 &K3III          &Lick/IDS std \\ 
HD061935 &G9III          &Lick/IDS std \\ 
HD066141 &K2III          &Lick/IDS std \\ 
HD071597 &K2III          &velocity std \\ 
HD083618 &K2.5III        &Lick/IDS std \\ 
HD088284 &K0III          &Lick/IDS std \\ 
HD095272 &K1III          &Lick/IDS std \\ 
HD219449 &K0III          &Lick/IDS std \\ 
HD221148 &K3III variable &Lick/IDS std \\ 
L745-46A &DF             &spec. std \cite{oke74} \\ 
GD~108   &sd:B           &spec. std \cite{oke90} \\ \hline 
    \end{tabular}
\end{table}

\subsection{Basic data reduction}
Most of the basic data reduction steps have been performed with
packages under {\tt IRAF}. For each night individually the science
frames were overscan corrected and a bias frame was subtracted. A few
bad columns were removed by linear interpolation. From several
domeflats and skyflats a final flatfield accounting for the
pixel-to-pixel variations and vignetting was constructed and applied to
the frames. Cosmic rays were removed using the {\tt cleanest} task in
the {\tt REDUCEME} package \cite{car98}. This task automatically
detects and removes cosmic rays via a sophisticated deviation
algorithm, while at the same time one can interactively inspect
potential cosmic rays in sensitive areas such as close to the galaxy
centre. The wavelength solution was determined from Th-Ar-lamp spectra
which were taken before and after most of the science observations. The
rms residual in the wavelength fit was typically 0.1~- 0.2~\AA. Finally
the sky was subtracted.

The central spectrum for each galaxy was extracted by fitting a low
order polynomial to the position of the centre along the wavelength
direction, re-sampling the data in the spatial direction and finally
co-adding the spectra within a 5 pixels aperture yielding an effective
aperture of 2\farcs3$\times$3\farcs85. Multiple exposures of the same
galaxy were combined. The resulting S/N in the spectra ranges from
$\sim 30$~[\AA$^{-1}$] for the faintest galaxies to more than
100~[\AA$^{-1}$] for the brightest ones (measured in a $\sim$100~\AA\/
wide region just bluewards of the \mgb\/ feature). For stars we used
the {\tt IRAF} task {\tt apall} to extract 1d-spectra. All galaxy and
stellar spectra were logarithmically rebinned to a common wavelength
range and increment. Finally the continuum shape of our spectra was
corrected to a relative flux scale with the help of the
spectrophotometric standard stars.

\subsection{Kinematics}
In order to correct the line-strength indices for velocity dispersion
broadening and to construct index--$\sigma_0$ relations we need to
measure the central velocity dispersion for each galaxy. Estimates were
derived with the Fourier correlation quotient (FCQ, version 8) method
\cite{ben90,ben94}. For the FCQ analysis the spectra were rebinned to
twice the original spectral sampling and a wavelength range of 4876 to
5653~\AA\/ was extracted. Note that the \Hb\/ feature is excluded from
the wavelength range as it proved to be a source of severe template
mismatch for galaxies with strong Balmer absorption. As we only
consider central spectra in this paper we fit a pure Gaussian profile
to the broadening function, neglecting higher order terms. To check the
reliability of the FCQ analysis we used eight different G \& K-giant
template stars. For galaxies with a central velocity dispersion of
$\sigma_0 \ge 70$~\kms, all eight template stars give very similar
results and an average value was adopted. The rms scatter between
different template stars is 0.007 in log units for galaxies with
$\sigma_0 \ge 100$~\kms. For galaxies with $70 \le \sigma_0 <
100$~\kms\/ the rms scatter increases to 0.024 and for galaxies with
$\sigma_0 < 70$~\kms\/ we find an rms scatter of 0.074. The uncertainty
introduced by different template stars was comparable or larger than
the internal error estimates of the FCQ program. Note, that for
galaxies with $\sigma_0 < 70$~\kms\/ some template stars gave a poor
fit to the broadening function and were excluded from the template
sample. Only remaining measurements were averaged.

Using this procedure, velocity dispersions as low as $\sim50$~\kms\/
could be recovered, although systematic errors will start to dominate
for $\sigma_0 < 90$~\kms. As our spectral resolution is rather low
compared to velocity dispersions of $\sim 50-60$~\kms\/ we emphasize
that for these faint galaxies our velocity dispersions are only rough
estimates. The final velocity dispersion errors for galaxies with
$\sigma \ge 70$~\kms\/ ($\Delta \log \sigma_0 = 0.022$) were derived by
a literature comparison (see Appendix, Figure~\ref{fig:comp_sig}). For
galaxies with $\sigma_0 < 70$~\kms\/ we adopt the mean rms scatter of
the template stars ($\Delta \log \sigma_0 = 0.074$).

\section{LICK/IDS CALIBRATION}
\label{sec:lick_cal}
The wavelength range of our spectra covers 16 different line-strength
indices, such as Mg$_2$, \Hb\/ and \HgA, in the Lick/IDS system which
is described in detail in Worthey (1994, hereafter W94), Worthey \&
Ottaviani (1997, hereafter WO97) and Trager et~al. (1998). In the
following analysis we use an updated version of the W94 models which is
available from Dr. G.~Worthey's home page. The updates affect only
models where $\rmn{[Fe/H]} \le -1.0.$ and are most noticable for the
\Hb\/ index.  For a recent study of the behaviour of the Balmer indices
at low metallicities see Poggianti \& Barbaro (1997) and Maraston,
Greggio \& Thomas (1999). Before one can compare the measured indices
with model predictions by \eg\/ W94 and Vazdekis et~al. (1996,
hereafter V96), the measurements have to be carefully calibrated to the
Lick/IDS system. Generally there are three effects to account for: (a)
the difference in the spectral resolution between the Lick/IDS system
and our set-up, (b) the internal velocity broadening of the observed
galaxies and (c) small systematic offsets caused by \eg\/ continuum
shape differences.

{\bf (a)} In order to account for diffrences in spectral resolution we
broadened the spectra with a Gaussian of wavelength dependent width,
such that the Lick/IDS resolution was best matched at each wavelength
(see Figure~7 in WO97). After this step our spectra should resemble
very well the general properties of the original spectra obtained by
the Lick group.

{\bf (b)} In a second step we need to correct the indices for velocity
dispersion broadening. The observed spectrum of a galaxy is the
convolution of the integrated spectrum of its stellar population(s) by
the instrumental broadening and the distribution of line-of-sight
velocities of the stars. These effects broaden the spectral features,
in general reducing the observed line-strength compared to the
intrinsic values. In order to compare the raw index measurements for
galaxies with model predictions we calibrate the indices to zero
velocity dispersion. Spectra of 15 different G9-K4 giant stars were
first broadened to the Lick/IDS resolution and then further broadened
using a Gaussian to $\sigma=20-360$~\kms\/ in steps of 20~\kms. The
indices are then measured for each star and $\sigma$-bin and a
correction factor, $C(\sigma)$, such that
$C(\sigma)=$Index(0)/Index($\sigma$) is determined.

Figure~\ref{fig:vel_disp_corr} in the Appendix shows the dependence of
the correction factor on $\sigma$ for all 16 indices. Note that for the
molecular indices Mg$_1$ and Mg$_2$ and the index \HgF\footnote{This
  index is actually not a molecular index but typical index values are
  close to zero hence a correction {\em factor} can degenerate.} the
correction factor is defined as
$C(\sigma)=$Index(0)~--~Index($\sigma$).  The scatter in $C(\sigma)$ at
360~\kms\/ was $< 5\%$ for all indices but \Hb.

It is worth looking in detail why the \Hb\/ velocity dispersion
correction seems to be so insecure. The derived correction factors are
only useful if the stars used for the simulations resemble the galaxy
spectra. In principle one might expect a dependence of the correction
factor on line-strength -- but most indices do not show such a
behaviour. In fact \Hb\/ is the only index where we find a significant
influence of line-strength on the correction factor at a given
$\sigma$. It turns out that stars which exhibit \Hb-absorption $\le
1.1$~\AA\/ lead to correction factors of $C(\sigma) < 1.0$ and stars
with \Hb-absorption $>1.1$~\AA\/ imply positive corrections. In the
Fornax sample there are no galaxies with \Hb\/ absorption line-strength
of less than 1.4~\AA, hence only stars with a \Hb\/ index greater
1.1~\AA\/ have been used to evaluate the correction factor (scatter $<
5\%$).

Another way to check the accuracy of the velocity dispersion
corrections is to use galaxy spectra with small internal velocity
dispersions as templates and treat them in the same way as stars. The
galaxies NGC1373, NGC1380A, NGC1336, IC1963 and ESO358-G59 were used
for this purpose. They span a range in \Hb\/ absorption of $\sim1.7 -
3$~\AA\/ and a range in central velocity dispersion of $\sigma_0 = 54 -
96$~\kms.  In Figure~\ref{fig:vel_disp_corr} the galaxies are
represented by open circles and they agree very well with the stellar
correction for most of the indices. As expected for \Hb, the galaxies
match the results from stars with a \Hb\/ absorption $>1.1$~\AA.

The final correction factors are derived by taking the mean of 15 stars
and the five galaxies in each $\sigma$-bin (solid line in
Figure~\ref{fig:vel_disp_corr}). The velocity dispersion corrections
are applied by a {\tt FORTRAN} program which reads in the raw
index-measurements from continuum corrected and resolution corrected
galaxy spectra. For each galaxy and index it applies a correction for
velocity dispersion. The program linearly interpolates between
$\sigma$-bins and also adds the error from the velocity-dispersion
correction factor to the raw Poisson error of the spectra. As the error
in the correction factor is much bigger than any error caused by
uncertainties in $\sigma$, we assumed the velocity dispersion of the
galaxies to be error free. 

{\bf (c)} Although we have matched very well the spectral resolution of
the Lick system, small systematic offsets of the indices introduced by
continuum shape differences are generally present (note that the
original Lick/IDS spectra are not flux calibrated). To establish these
offsets we compared our measurements for stars in common with the
Lick/IDS stellar library. In total we observed 13 different Lick/IDS
stars. Figure~\ref{fig:lick_off} in the Appendix shows the difference
between Lick/IDS measurements and ours {\em after} the mean offset has
been removed. The mean offsets and associated errors for each index are
summarized in Table~\ref{tab:lick_off}. The star HD221148 was excluded
from the offset analysis because our index measurements proved to be
very different from the original Lick/IDS measurements -- possibly due
to its variable nature (see Table~\ref{tab:log_stars}). The formal
error in the offset is evaluated by the mean standard deviation of
stars with respect to the mean offset divided by $\sqrt{n_{stars}-1}$.

\begin{table}
  \caption[]{Lick/IDS Offsets }
  \label{tab:lick_off}
  \begin{tabular}{ll} \hline 
     Index    & offset (Lick/IDS - AAT) \\ \hline 
     G4300    & $+0.21\pm0.09$   \AA  \\     
     Fe4383   & $+0.60\pm0.13$   \AA  \\     
     Ca4455   & $+0.37\pm0.06$   \AA  \\     
     Fe4531   & $+0.00\pm0.10$   \AA  \\     
     C$_2$4668& $-0.19\pm0.17$   \AA  \\     
     H$\beta$ & $-0.05\pm0.04$   \AA  \\
     Fe5015   & $+0.00\pm0.08$   \AA  \\     
     Mg$_1$   & $+0.003\pm0.002$ mag  \\    
     Mg$_2$   & $+0.023\pm0.003$ mag  \\
     \mgb     & $+0.15\pm0.09$   \AA  \\  
     Fe5270   & $+0.07\pm0.05$   \AA  \\ 
     Fe5335   & $+0.00\pm0.08$   \AA  \\ 
     Fe5406   & $+0.00\pm0.04$   \AA  \\
     Fe5709   & $+0.00\pm0.06$   \AA  \\
     \HgA     & $+0.45\pm0.28$   \AA  \\
     \HgF     & $+0.00\pm0.14$   \AA  \\ \hline
    \end{tabular}
\end{table}

Most off the indices show small offsets to the Lick/IDS system, similar
to the ones quoted in (WO97, Table 9). The rather large offset in
Mg$_2$ is due to a well known difference in continuum shape.

Recently Trager~et~al. (1998) published the Lick/IDS library of
extragalactic objects including 7 galaxies in the Fornax cluster and
NGC3379. We can check our previous offset evaluation by comparing our
galaxy measurements with Trager et~al. For this purpose we extracted a
3 pixels central aperture (2\farcs3$\times$2\farcs44) for our galaxies
matching the Lick/IDS standard aperture of 1\farcs4$\times$4\arcsec.
Our indices are then corrected for velocity dispersion as described in
paragraph (b) and the offsets from Table~\ref{tab:lick_off} are
applied. The results are overplotted in Figure~\ref{fig:lick_off} in
the Appendix (filled symbols). The galaxies show for all indices more
scatter around the mean offset than the stars which is somewhat
reflected in the bigger error bars, but there are also some outliers.
This is not surprising as seeing effects and aperture differences will
introduce some non-reproduceable offsets for individual galaxies.
Furthermore we note that the Lick group had to observe the Fornax
galaxies at a very high airmass. With the possible exception of the
indices G4300 and Fe4383 the offset inferred from the galaxy comparison
are consistent with the stellar comparison.

The offsets listed in Table~\ref{tab:lick_off} were applied to all
indices after the correction for velocity dispersion. Note that the
Lick/IDS-system offset-error is a constant value and does not depend on
the velocity dispersion of the galaxy itself. Therefore we did not
include this error in the individual index errors but rather quote for
each index a common offset error (see also Table~\ref{tab:lick_off}).
The final corrected central (2\farcs$\times$3\farcs85) index
measurements and associated errors for the Fornax galaxies and NGC3379
are presented in Table~D2 in the Appendix. For each galaxy we give the
index measurement in the first row and the 1$\sigma$ error in the
second row.

Note that for the galaxies NGC1381 and NGC1427 we combined three
exposures yielding a very high S/N spectrum. Here our index-error
estimation taking into account only the Poisson error becomes invalid
because of other error sources such as the wavelength calibration,
continuum correction and aperture effects. By comparing individual
exposures we established that 1.5 $\times$ the original Poisson error
estimate is a good indicator of the true error. This adjusted error was
adopted in Table~D2 and for any further analysis.

\section{A consistency test of the model predictions}
\label{sec:consistent}
\begin{figure}
\epsfig{file=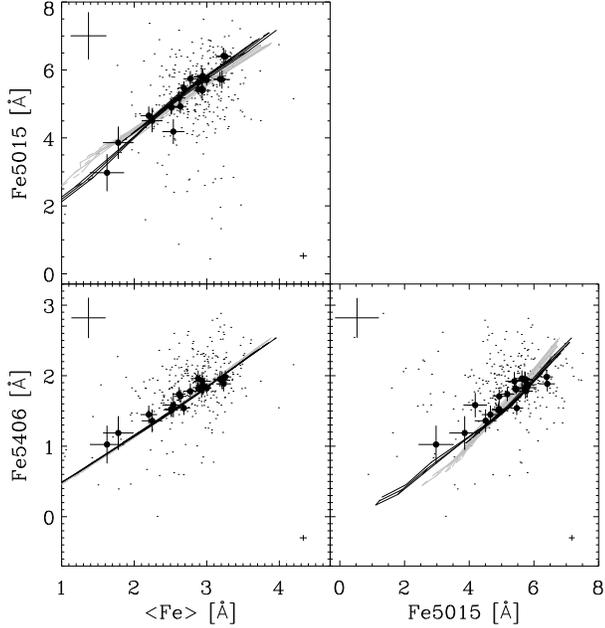,height=8.5cm,width=8.5cm}
\caption[]{\label{fig:ind_ind_fe}Index {\em vs}\/ index plots
  for three well established Fe-indicators in the Lick/IDS system. The
  filled circles and small dots represent AAT and Lick/IDS galaxy
  measurements respectively. The error bar in the upper left corner
  represents the average observational error for Lick/IDS galaxies
  whereas for the AAT data the observational errors are shown for each
  individual galaxy. The error bar in the lower right corner shows the
  rms error in the offset to the Lick/IDS system for the AAT data.
  Overplotted are model predictions by Worthey (1994, black lines) and
  Vazdekis et~al. (1996, grey lines). Note that Worthey models use a
  Salpeter IMF whereas Vazdekis models use a bimodal IMF which is very
  similar to Salpeter for $\rmn{M} > 0.6~\rmn{M}_{\odot}$.}
\end{figure}

For the following analysis of the nuclear stellar populations
(Section~\ref{sec:nucstell}) it is extremely important that our index
measurements are accurately calibrated onto the Lick/IDS system which
is based on the Lick/IDS stellar library \cite{wor94b}. Here we
investigate the accuracy and consistency of our calibration and the
model predictions by presenting index--index plots which are almost
degenerate in age and metallicity.  In that way the model predictions
cover only a small ``band'' of the parameter space and they should
trace the relation of the galaxies if the models describe accurately
the galaxy properties and our calibration is accurate.
Figure~\ref{fig:ind_ind_fe} shows the relation between
$<$Fe$>$\footnote{$<$Fe$>$=(Fe5270+Fe5335)/2}, Fe5015 and Fe5406 for
our sample of galaxies (filled circles) and the original Lick/IDS
sample of galaxies (Trager et~al. 1998, small dots).  Overplotted are
models by W94 (black lines) and V96 (grey lines). The plots show a good
agreement between index measurements and the model predictions. The
reduced Poisson noise of our data set compared to the Lick/IDS
measurements can be clearly seen (see also Figure caption). We note
that the model predictions of Vazdekis and Worthey are in good
agreement.


\begin{figure}
\epsfig{file=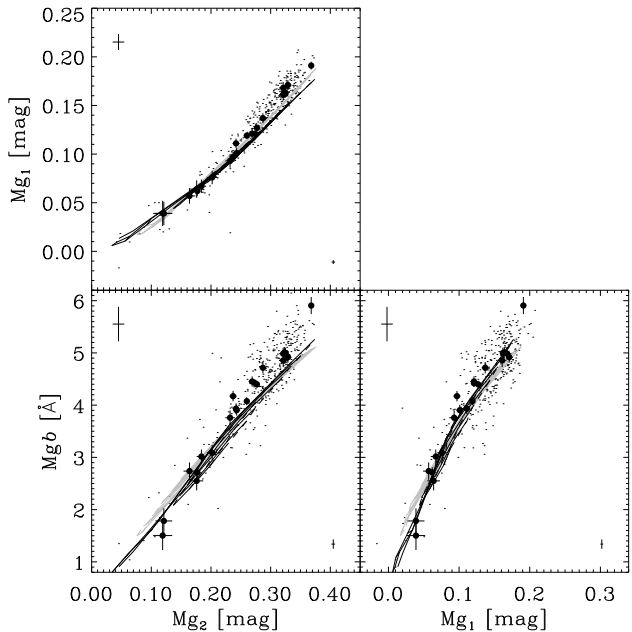,height=8.5cm,width=8.5cm}
\caption[]{\label{fig:ind_ind_mg}Index {\em vs}\/ index plots
  for the three Mg indices in the Lick/IDS system. Overplotted are
  model predictions by Worthey (1994, black lines) and Vazdekis et~al.
  (1996, grey lines).}
\end{figure}

\begin{figure}
\epsfig{file=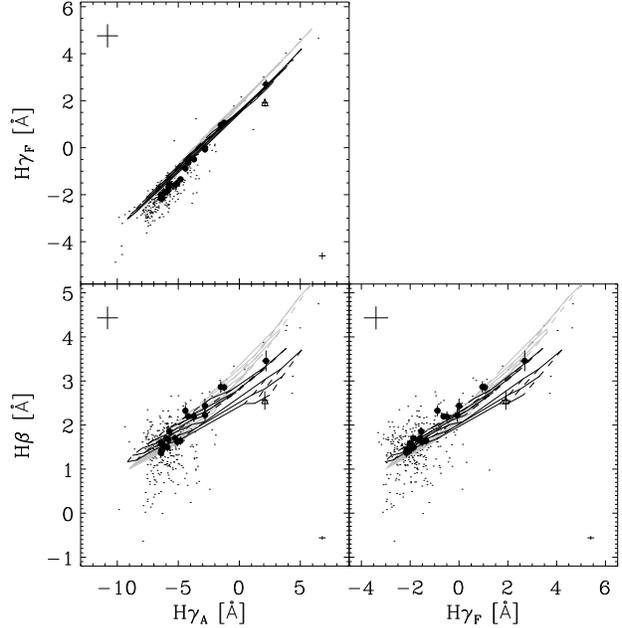,height=8.5cm,width=8.5cm}
\caption[]{\label{fig:ind_ind_h}Index {\em vs}\/ index plots
  for three Balmer line indices in the Lick/IDS system. Symbol
  definitions are the same as in Figure~\ref{fig:ind_ind_fe}, except
  for the galaxy ESO358-G25 which is represented by an open triangle.
  This galaxy is affected by emission in the Balmer lines. Overplotted
  are model predictions by Worthey (1994, black lines), Worthey \&
  Ottaviani (1997, black lines) and Vazdekis et~al. (1996, grey
  lines).}

\end{figure}

A similar analysis of the three Mg indices is shown in
Figure~\ref{fig:ind_ind_mg}. Here we find a significant deviation of
the measured index values compared to the model predictions for metal
rich and/or old stellar populations. The deviations are seen in the
Fornax sample as well as in the original Lick/IDS galaxy spectra (see
also Worthey 1992, Figure~5.12~\& 5.13). We therefore note that this
discrepancy is inherent to the Lick/IDS system \& models and any models
which use the Lick/IDS fitting functions are likely to show the same
offset. In Figure~\ref{fig:ind_ind_h} we present the Balmer indices
\Hb, \HgA\/ and \HgF. Here we find generally good agreement with small
deviations between model predictions and data at low values of \HgA\/
{\em vs}\/ \HgF\/ which is present in our data and the original
Lick/IDS measurements. 

Figures~\ref{fig:ind_ind_fe} to \ref{fig:ind_ind_h} suggest that our
Lick/IDS calibration is very consistent with the original galaxy
measurements of the Lick group. However, we note that small, systematic
offsets exist between the parameter space covered by galaxies and the
model predictions for Magnesium at high index values and for Balmer
lines at low index values.

\section{THE NUCLEAR STELLAR POPULATIONS}
\label{sec:nucstell}
The aim of this section is to derive estimates of the mean (luminosity
weighted) ages and metal abundances of early-type galaxies in the
Fornax cluster. As pointed out by W94, the determination of the ages
and metallicities of old stellar populations is complicated by the
similar effects that age and metallicity have on the integrated
spectral energy distributions. However, this degeneracy can be
partially broken by plotting a particular age sensitive index, such as
one of the Balmer line indices, against a more metallicity sensitive
index. The usefulness of this approach has been demonstrated by many
authors \cite{gon93,fis95,meh98,kun98a,jor99}. However, as we will see
in this Section, among other issues the treatment of non-solar
abundance ratios is a crucial parameter in the determination of
absolute age and metallicity estimates. We will also investigate the
effects of nebular emission and composite stellar populations on
age/metallicity estimates in Sections~\ref{sec:emission} \&
\ref{sec:composite} respectively, before we present our best
age/metallicity estimates of the Fornax early-type galaxies in
Section~\ref{sec:bestage}.

\subsection{Non-solar abundance ratios}
\label{sec:non_solar}
%
%

\begin{figure*}

\epsfig{file=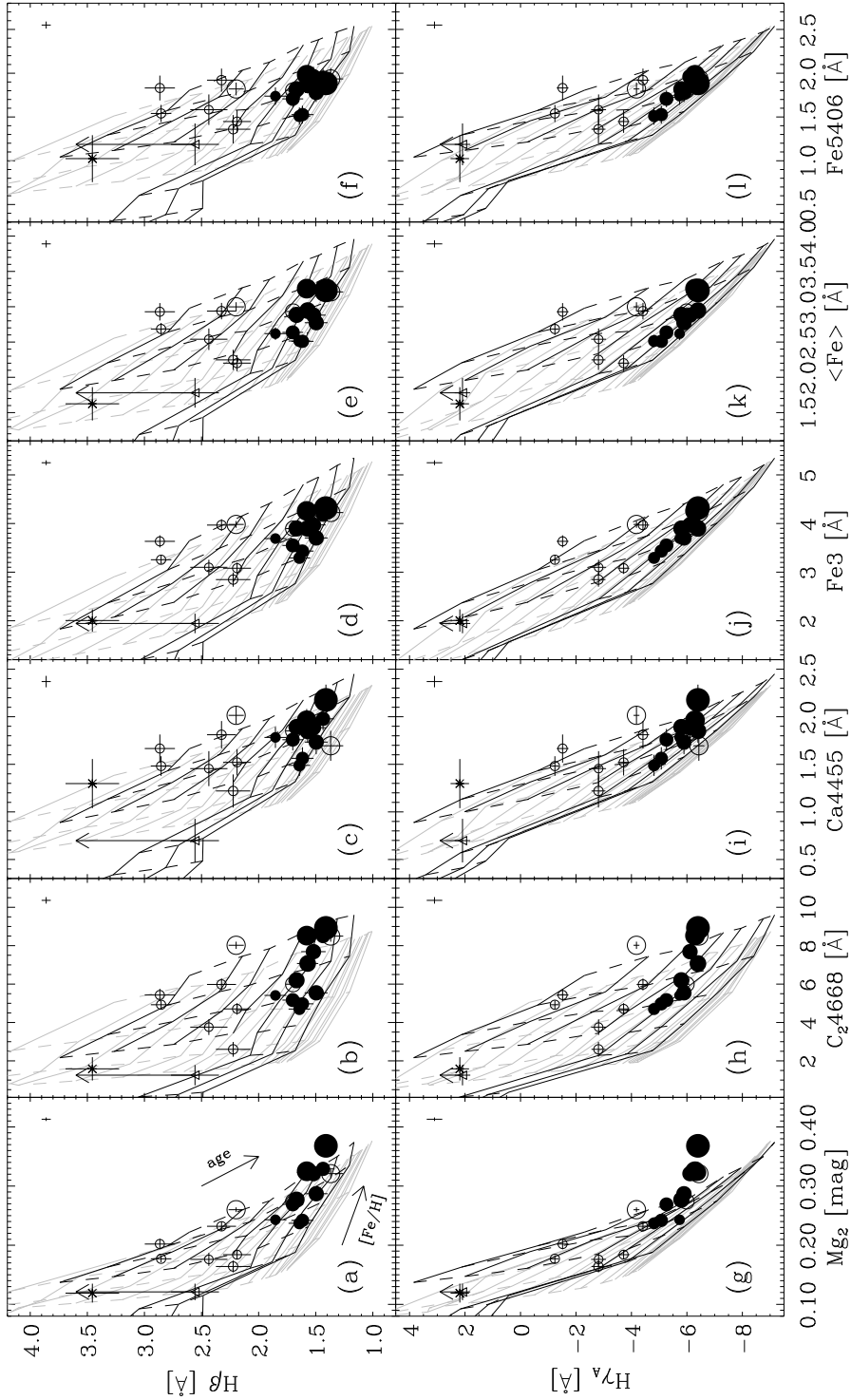,angle=90,height=21cm}

\caption[]{\label{fig:hb_metal}The two age sensitive indices \Hb\/ and
  \HgA\/ plotted against six metallicity indicators: Mg$_2$, C$_2$4668,
  Ca4455, Fe3, $<$Fe$>$ and Fe5406. Models by Worthey (1994, black
  lines), Worthey \& Ottaviani (1997, black lines) and Vazdekis et~al.
  (1996, grey lines) are overplotted. The solid lines represent isoage
  lines whereas the dashed lines are lines of constant metallicity.
  Filled circles and open circles represent ellipticals and S0s
  respectively. The star and open triangle represent possible
  post-starburst and starburst galaxies respectively. The arrow
  attached to ESO358-G25 (open triangle) indicates an emission
  correction. The symbol size is scaled with the central velocity
  dispersion of the galaxies. The cross in the upper right corner of
  each panel indicates the rms uncertainty in the transformation to the
  Lick/IDS system.  Observational errors are plotted as individual
  error bars on the data points.}

\end{figure*}

In Figure~\ref{fig:hb_metal} we present age/metallicity diagnostic
diagrams of six metallicity sensitive indices (Mg$_2$, C$_2$4668,
Ca4455, Fe3, $<$Fe$>$ \& Fe5406) plotted against the age sensitive
Balmer line indices \Hb\/ and \HgA\/ (the new index Fe3 is defined in
Equation~\ref{eq:fe3}). Figure~\ref{fig:hb_metal}h is a reproduction
from Kuntschner \& Davies (1998) with minor data up-dates.  Overplotted
are model predictions from W94, WO97 (black lines) and V96 (grey
lines). Solid lines represent lines of constant age and the dashed
lines are lines of constant metallicity.  The Worthey models span a
range in age of 1.5--5 Gyr with [Fe/H]=-0.225 to 0.5 and 8--17 Gyr with
[Fe/H]=-2 to 0.5. The V96 models span a range in age of 1--17.4 Gyr
with [Fe/H]=-0.7 to 0.4. The direction of increasing age and
metallicity is indicated in Figure~\ref{fig:hb_metal}a by arrows.

Our previous result from Kuntschner \& Davies (1998), that Fornax
ellipticals form a sequence in metallicity at old ages and that the S0s
spread to younger ages is confirmed in all diagrams. Examining the
diagrams in detail one can see that the mean age and metallicity of the
sample changes from diagram to diagram; \eg\/ the ellipticals appear
older and more metal poor in the $<$Fe$>$ {\em vs}\/ \HgA\/ diagrams
compared to the Mg$_2$ {\em vs}\/ \HgA\/ diagram. This effect was
previously reported and recently reviewed by Worthey (1998). It is now
widely accepted that this discrepancy in the model predictions is
caused by non-solar abundance ratio effects. For example Mg as measured
by the Mg$_2$ index is overabundant compared to Fe in luminous
elliptical galaxies, \ie\/ {${\rmn [Mg/Fe]} > 0$}
\cite{oco76,pel89,wor92,dav93,wei95,jor97,jor99}.

\begin{figure}
\epsfig{file=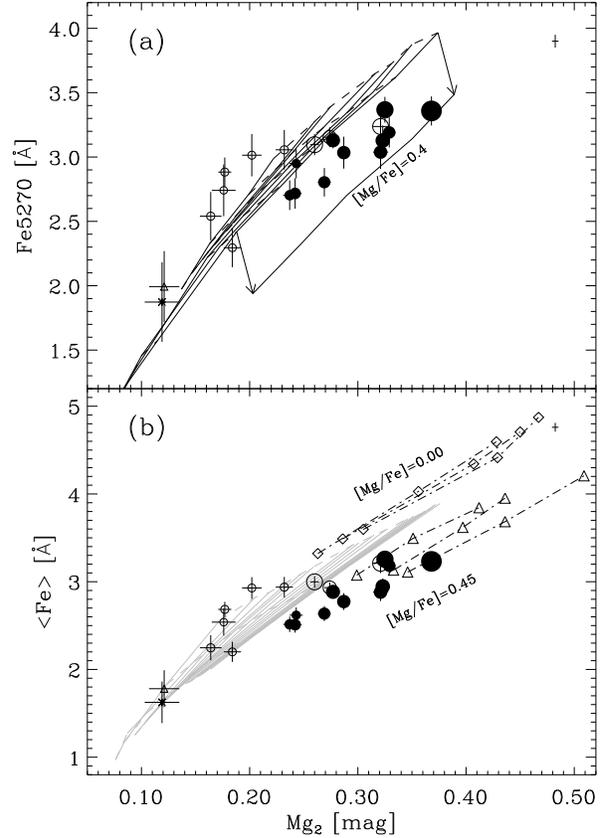,width=8.5cm}
\caption[]{\label{fig:mg_fe}(a) Mg$_2$ {\em vs}\/ Fe5270 equivalent width 
  diagram for the complete sample of Fornax early-type galaxies.
  Overplotted are models by Worthey (1994) and a correction for
  [Mg/Fe]=0.4 for the 17 Gyrs isoage line (taken from Greggio 1997).
  (b) Mg$_2$ {\em vs}\/ $<$Fe$>$ diagram. Overplotted are models by
  Vazdekis (1996, grey lines) and two models by Weiss, Peletier \&
  Matteucci (1995, dot-dashed lines). The Weiss et~al. models are
  calculated for three ages, 12, 15 \& 18 Gyrs (dot-dashed lines
  represent lines of constant age) at Z=0.02, 0.04 \& 0.07, a mixing
  length parameter $\alpha_{\rmn MLT} = 1.5$ and somewhat different
  mixes of heavy elements. Steps in metallicity are shown as diamonds
  for ${\rmn [Mg/Fe]} = 0.0$ and as triangles for ${\rmn [Mg/Fe]} =
  0.45$.  Symbol definitions as in Figure~\ref{fig:hb_metal}.}
\end{figure}

The Mg overabundance can be examined in a Mg-index {\em vs}\/ Fe-index
plot \cite{wor92}. In such a diagram the model predictions cover only a
narrow band in the parameter space as effects of age and metallicity
are degenerate. Figure~\ref{fig:mg_fe} shows plots of $<$Fe$>$ \&
Fe5270 {\em vs}\/ Mg$_2$ for the Fornax sample. Overplotted are model
predictions from W94, V96 and Weiss et~al. (1995). We assume that the
models reflect solar abundance ratios if not stated otherwise, \ie\/
${\rmn [Mg/Fe]} = 0$. If the model predictions accurately resemble the
galaxy properties they should trace the observed relation. The measured
line-strength of most of the S0s agrees with the model predictions,
perhaps 3--4 galaxies having slightly low Mg$_2$ absorption compared to
Fe5270 \& $<$Fe$>$. However, for most of the ellipticals and the S0
NGC1380, the models predict too little Mg-absorption at a given
Fe-absorption strength. Additionally the most metal rich galaxies are
the furthest away from the model grids. Using the Mg-overabundance
correction by Greggio (1997, see Figure~\ref{fig:mg_fe}a) and the
models for [Mg/Fe]=0.45 by Weiss et~al. (1995), we conclude that the
stellar populations of Fornax ellipticals and the bulge of NGC1380 are
Mg-overabundant relative compared to Fe. The overabundance ranges
between [Mg/Fe]=0.0 to $\sim$0.4. We note that there exists
considerable spread in overabundance at a given Fe-line strength in our
sample.

Non-solar abundance ratios do exist not only in elliptical galaxies but
also in our own galaxy where stars show an overabundance for
$\alpha$-elements\footnote{$\alpha$ includes the elements O, Mg, Si, S,
  Ca, and Ti} at [Fe/H]$\la$0.0 \cite{edv93,mcw97}. Of course if those
stars are incorporated in a stellar library which in turn is used for
model predictions of integrated stellar populations, the predictions
will be somewhat $\alpha$-element overabundant at low metallicities. We
therefore note that models which use the Lick/IDS fitting functions are
probably $\alpha$-element overabundant at low metallicities which makes
it more difficult to interpret trends in diagrams such as
Figure~\ref{fig:mg_fe}.

\begin{figure}
\epsfig{file=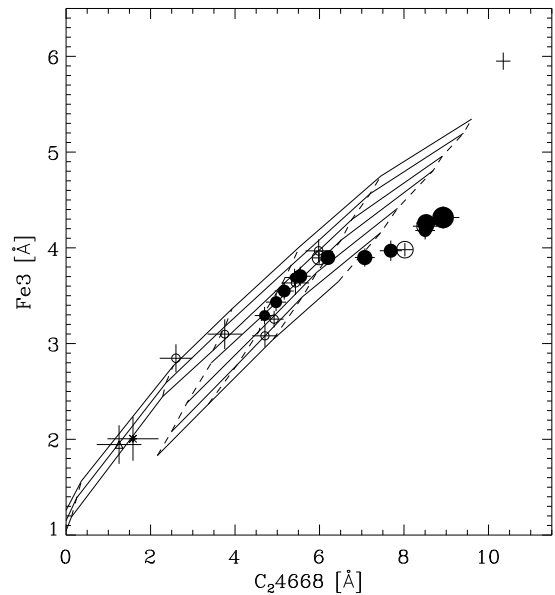,width=8.5cm}
\caption[]{\label{fig:c4668_fe}C$_2$4668 equivalent width {\em vs}\/ Fe3
  equivalent width. Overplotted are models by Worthey (1994). The
  C$_2$4668 index shows evidence of overabundance compared to Fe3 at
  strong absorption strength ($>$ 6.5~\AA). The symbol size is scaled
  with the central velocity dispersion of the galaxies.}
\end{figure}

Several indices covered by our wavelength range show deviations from
the model predictions when compared to the average Fe index: Mg$_1$,
Mg$_2$, \mgb, Fe5709 \& C$_2$4668. Fe5709 is a very weak index and its
correction for velocity dispersion broadening may well be insecure, so
we cannot draw any firm conclusions. C$_2$4668 is an important index
because it shows the strongest total metallicity sensitivity in the
Lick/IDS system \cite{wor98} and is therefore preferentially used in
age/metallicity diagnostic diagrams. In Figure~\ref{fig:c4668_fe} we
present a plot of C$_2$4668 {\em vs}\/ Fe3. Fe3 is a combination of
three prominent Fe lines, thus maximizing its sensitivity to Fe while
minimizing the Poisson errors:

\begin{equation}
  \label{eq:fe3}
  {\rmn Fe3} = \frac{ {\rmn Fe4383} + {\rmn Fe5270} + {\rmn Fe5335} }{3}
\end{equation}

As a consequence of the extreme metallicity sensitivity of C$_2$4668,
the models are not as degenerate as in the previous plots. Nevertheless
it is clear that for a C$_2$4668 absorption strength in excess of
$\sim$6~\AA\/ the model predictions do not follow the observed trend
(see also Kuntschner 1998). Hence, we conclude that C$_2$4668, or
better at least one of the species contributing to the index, is
overabundant compared to Fe in metal rich Fornax galaxies. Can this
overabundance be caused by Mg as seen in the Fe {\em vs}\/ Mg plot
(Figure~\ref{fig:mg_fe})? Due to the proximity of metal absorption
lines in the optical wavelength region non of the Lick/IDS indices
measures the abundance of only a particular element such as Fe or Mg.
There are always contributions from other elements or molecules to an
index \cite{tri95}. In particular, the C$_2$4668 index has a relatively
wide central bandpass (86.25~\AA) including a wide range of metal
lines. The most dominant species here is carbon in form of C$_2$-bands
which blanket the central bandpass \cite{tri95}. Yet, more important
here is that according to Tripicco~\& Bell (1995; Table 6, cool giants)
the C$_2$4668 index {\em decreases} when the Mg-abundance (or
Oxygen-abundance) is {\em increased} (at fixed abundances of all other
elements). Therefore the 'overabundance' of C$_2$4668 cannot be caused
by Mg. What exactly drives the overabundance of C$_2$4668 compared to
Fe remains to be seen.

The overabundance of certain elements compared to Fe has a profound
effect on the use of age/metallicity diagnostic diagrams if the model
predictions reflect only solar abundance ratios. Recalling
Figure~\ref{fig:hb_metal}f (see also Kuntschner \& Davies 1998) we find
that not only are the metallicities (measured as [Fe/H]) overestimated
due to the C$_2$4668 overabundance but furthermore much of the age
upturn at high metallicities is caused by the overabundance. The effect
of changing the age estimates is caused by the residual age/metallicity
degeneracy which is still present in all diagrams in
Figure~\ref{fig:hb_metal}.  Only if the index-combination breaks the
degeneracy completely, \ie\/ if lines of constant age and constant
metallicity are perpendicular, would the age estimates not be affected
by non-solar abundance ratios.  We further note that trends in
abundance ratios within a dataset such as our Fornax sample (increasing
Mg/Fe with galaxy mass) can lead to artificial relative age trends in
diagrams such as in Figure~\ref{fig:hb_metal}f. Taking into account not
only the uncertainties introduced by non-solar abundance ratios but
also other model parameters such as which isochrone library to use, it
seems very insecure to derive {\em absolute} age estimates from the
currently available stellar population models.

Introducing non-solar abundance ratios in model predictions is rather
complicated, as accurate model predictions do need a stellar library
covering the whole parameter space of T$_e$, log g, [Fe/H] {\em and}\/
[Mg/Fe]. Furthermore, new isochrone calculations may well be needed for
each [Mg/Fe] bin: recently Salaris \& Weiss (1998) suggested that
scaled-solar isochrones cannot be used to replace Mg-enhanced ones at
the same total metallicity. The latter will not only change the model
predictions for indices such as Mg$_2$ but may affect all indices and
in particular the age sensitive ones such as \Hb\/ and \HgA\/ (see also
Worthey 1998). However, note that Weiss, Peletier \& Matteucci (1995)
concluded in their study that scaled solar isochrones are sufficient to
calculate model predictions for non-solar abundance ratios.

\begin{figure*}
\epsfig{file=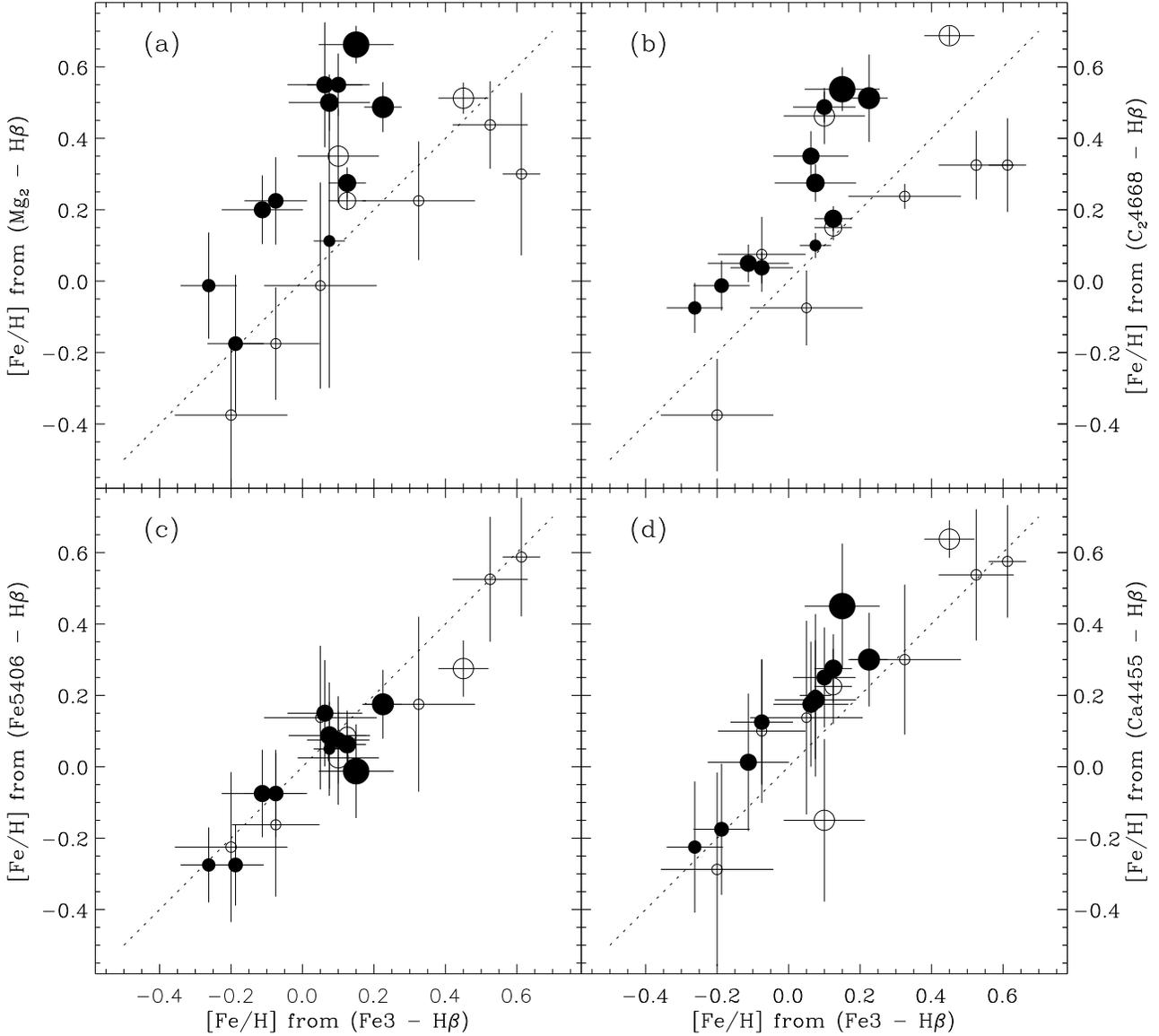,width=18cm}
\caption[]{\label{fig:overab}Metallicity estimates derived from four
  age-metallicity diagnostic diagrams all using \Hb\/ as age indicator
  but different metal lines (Mg$_2$, Fe5406, C$_2$4668 \& Ca4455) are
  compared with the metallicity estimates from the Fe3 {\em vs}\/ \Hb\/
  diagram. The filled circles represent elliptical galaxies and the
  open circles stand for the S0s. The symbol size is scaled with the
  central velocity dispersion of the galaxies.}
\end{figure*}

Another way to examine non-solar abundance ratios is to compare the
metallicity estimates derived from different metal lines using the same
age indicator. Figure~\ref{fig:overab} compares the metallicity
estimates taken from Mg$_2$, C$_2$4668, Fe5406, Ca4455 {\em vs}\/ \Hb\/
diagrams with the estimates taken from a Fe3 {\em vs}\/ \Hb\/ diagram.
Here Fe3 serves as our mean Fe-abundance indicator. The metallicity
estimates are derived from the V96 models\footnote{These models have a
  bimodal IMF which is very similar to Salpeter for $\rmn{M} >
  0.6~\rmn{M}_{\odot}$. Note that for an age of $\sim$17 Gyrs V96
  models predict $0.1 - 0.2$~\AA\/ less \Hb-absorption compared to W94
  models. This of course will affect the absolute age estimates but has
  little affect on the metallicity estimates.}. To get more accurate
estimates, the age/metallicity--grid was expanded to a step size of
0.025 in [Fe/H] by linear interpolation. Furthermore the diagram was
extrapolated to [Fe/H]=+0.7 by linear extrapolation.  The age range of
1 to 17.4 Gyrs is covered by 18 grid points. Errors on the metallicity
estimates were derived by adding and subtracting the index error for
each galaxy individually (Poisson error and Lick/IDS offset error added
in quadrature) and re-deriving the metallicity estimates. The final
uncertainty displayed in Figure~\ref{fig:overab} was taken to be 0.7
times the maximum change in [Fe/H].

In panel (a) of Figure~\ref{fig:overab} we can clearly see that for
elliptical galaxies Mg$_2$ gives metallicity estimates which are larger
than those derived from Fe3 and there is a trend that the
Mg-overabundance increases with increasing metallicity. Most of the S0s
are consistent with solar or slightly less than solar abundance ratios
of Mg/Fe. However, the (more luminous) S0s NGC1380 \& NGC1381 show a
weak overabundance of Mg. The index C$_2$4668 (panel b) gives on
average high metallicity estimates compared to Fe3. Although three
galaxies with just above solar metallicity show solar abundance ratios.
As expected the Fe-index Fe5406 (panel c) is in good agreement with the
estimates derived from Fe3. The Ca4455 index (panel d) gives marginally
higher metallicity estimates compared to the Fe3 indicator. We note
that the Ca4455 index is more sensitive to a mix of heavy elements than
to Calcium on its own despite its name \cite{tri95}.

In conclusion we can confirm our previous results that Mg and C$_2$4668
are overabundant compared to Fe. The Mg-overabundance follows a trend
where metal rich (and luminous) Es show a stronger overabundance than
less luminous and metal poor galaxies.

\subsection{Nebular emission in early-type galaxies}   
\label{sec:emission}
So far we have concentrated on breaking the age/metallicity degeneracy
and the treatment of non-solar abundance ratios. A further important
issue when estimating ages and metallicities from line strength indices
is nebular emission. Elliptical galaxies normally contain much less
dust and ionized gas than spirals, in fact, for a long time they were
regarded as dust and gas free. However, spectroscopic surveys of large
samples of early-type galaxies revealed that about 50-60\% of the
galaxies show weak optical emission lines \cite{phi86,cal84}.
Typically the reported strength of emission lines such as [O{\small
  II}], [H$\alpha$] and [N{\small II}] $\lambda 6584$ indicates the
presence of only $10^3 - 10^5 M_{\odot}$ of warm ionized gas in the
centre. A more recent study of 56 bright elliptical galaxies by
Goudfrooij et~al. (1994) detected ionized gas in 57\% of their sample
and confirms the amount of ionized gas present. Additionally, HST
images of nearby bright early-type galaxies revealed that approximately
70-80\% show dust features in the nucleus \cite{dok95}. Stellar
absorption line-strength measurements can be severely affected if there
is emission present in the galaxy which weakens the stellar absorption
\cite{gou96}. For example, nebular \Hb-emission on top of the
integrated stellar \Hb-absorption weakens the \Hb-index and leads
therefore to wrong, \ie\/ too old age estimates.

The spectrum of ESO358-G25 shows clear emission in \Hb\/ and H$\gamma$
along with weak [O\,{\small III}] emission (see Kuntschner \& Davies
1998, Figure~3). As a consequence, the age is overestimated in
Figures~\ref{fig:hb_metal}~(a)--(f). The arrow attached to ESO358-G25
indicates a rough emission correction. However, it is extremely
difficult to accurately correct the \Hb-index in individual galaxies
for emission contamination. A much better method to reduce emission
contamination is to use higher order Balmer lines such as \Hg\/ as they
are less affected by nebular emission \cite{ost89}. Indeed, in
Figures~\ref{fig:hb_metal}~(g)--(l) the galaxy ESO358-G25 moves to much
younger ages. As none of the other galaxies move significantly to
younger ages we conclude that nebular emission is not very prominent in
our Fornax sample. This is supported by the absence of strong [O{\small
  III}]$\lambda 5007$ emission. Only 5 galaxies show emission above our
detection limit of $\sim$0.2~\AA. The strongest emission is detected in
ESO358-G25 with 0.7~\AA\/ equivalent width (for details see Kuntschner
1998).

\subsection{Effects of composite stellar populations}
\label{sec:composite}
\begin{figure*}
\epsfig{file=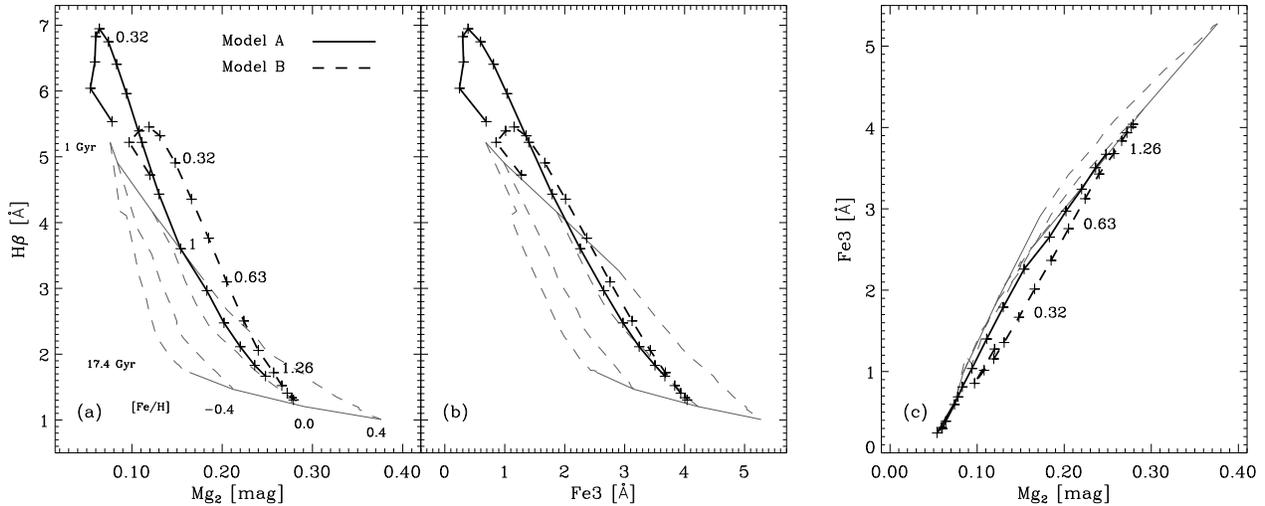,width=18cm}
\caption[]{\label{fig:comp_pos}Evolutionary tracks based on
  Vazdekis et~al. (1996) are shown in three index--index diagrams for a
  composite (Model A) of a 15~Gyrs old (solar metallicity, 90\% mass)
  and a young stellar population (solar metallicity, 10\% mass) at
  burst ages 0.1, 0.13, 0.16, 0.2, 0.25, 0.32, 0.4, 0.5, 0.63, 0.79,
  1.0, 1.26, 1.58, 2.0, 2.51 \& 3.16~Gyrs. The plus symbols along the
  tracks indicate the time steps. Age steps of 0.32 and 1~Gyrs are also
  indicated by numbers in panel (a). Model B represents a burst of 1\%
  (in mass) strength with the same metallicities and age steps as in
  Model A. Age steps of 0.32, 0.63 and 1.26~Gyrs are indicated in panel
  (a) \& (c). The region of normal SSP models (Vazdekis et~al. 1996) is
  shown as thin lines.}
\end{figure*}

Most of the S0s in our sample have luminosity weighted young stellar
populations with some of them having also high metallicities when
compared to single-burst stellar population (SSP) models. However,
these galaxies show only a central young stellar population on top of
an underlying older one as opposed to be entirely young. It is not
straight forward to compare {\em composite}\/ stellar populations with
SSP models \cite{djo97}. So, how reliable are the age, metallicity and
abundance ratio estimates taken directly from SSP models for these
young galaxies? In order to explore this issue we calculated model
predictions based on V96 for simple composite stellar populations. Two
representative tracks are shown in Figure~\ref{fig:comp_pos} for two
age/metallicity diagnostic diagrams and a plot of Fe3 {\em vs}\/ Mg$_2$
in order to explore the behaviour of abundance ratios: Model~A is a
15~Gyr old (90\% in mass) stellar population plus a burst (10\% in
mass) of varying ages from 0.1 Gyr to $\sim3$~Gyrs. Both populations
have solar metallicity.  For Model~B we reduced the burst fraction to
1\% (in mass) while the other parameters are the same as in Model~A.

Overall one can see in the age/metallicity diagnostic diagrams that for
a short time the burst population will dominate the integrated light
leading to strong \Hb\/ absorption and weak metal-line absorption. Then
the underlying old population becomes more and more important and after
$\sim$3~Gyrs the galaxy is almost back to its original place in the
diagram. However, the burst strength influences the exact track which
the galaxy takes in the diagram. For a burst of 10\% or 20\% (not
shown) in mass the tracks follow roughly the solar metallicity line in
the normal SSP models. Yet, for a small burst (1\% in mass) the
integrated light looks for a short while as having metallicities well
above solar. This effect is more pronounced for Mg$_2$ than for Fe3. Of
course, this in turn leeds to an artificially created overabundance
when these galaxies are compared to SSP models (see
Figure~\ref{fig:comp_pos}c). For bursts stronger than a few percent the
abundance ratios are not significantly affected.

In summary we find that composite stellar populations and in particular
small (in mass) bursts, such as used in our simple models, can lead to
an overestimation of the metallicity in the context of SSP models.
Abundance ratios can be affected in the sense that the Mg/Fe ratio is
too strong. Our model calculations show that these conclusions
qualitatively hold if the metallicity is changed or different
metallicities are combined. A more thorough investigation of these
issues would be very valuable but is beyond the scope of this paper
(see Hau, Carter \& Balcells (1999) for a more detailed analysis).

\subsection{Best age and metallicity estimates}
\label{sec:bestage}
\begin{figure}
\epsfig{file=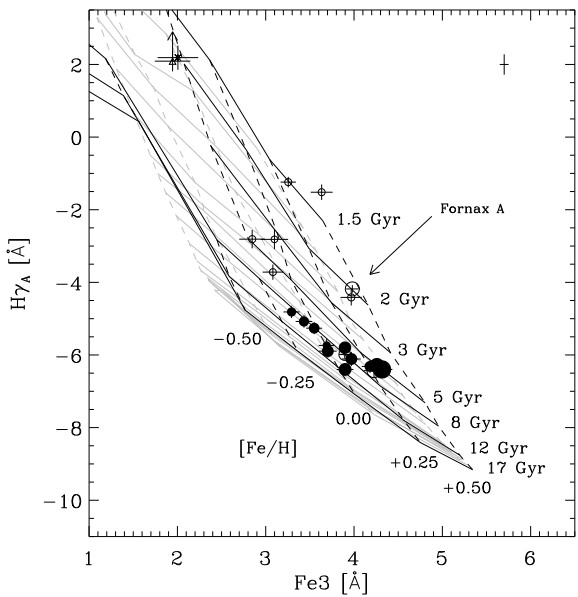,width=8.5cm}
\caption[]{\label{fig:fe3_hga}Fe3 equivalent width {\em vs}\/ \HgA\
  equivalent width. Filled circles and open circles represent
  ellipticals and S0s respectively. The star and open triangle
  represent possible post-starburst and starburst galaxies
  respectively. The cross in the upper right corner of each panel
  indicates the rms uncertainty in the transformation to the Lick/IDS
  system. The symbol size is scaled with the central velocity
  dispersion of the galaxies. Note that the two bright S0s are somewhat
  hidden in the sequence of Es.}
\end{figure}

Having examined some of the fundamental problems with applying stellar
population model predictions to observed line-strength indices we
present in Figure~\ref{fig:fe3_hga} what we consider our best
age/metallicity diagnostic diagram. A mean Fe-index (Fe3) is plotted
against an emission robust higher order Balmer line (\HgA). Due to the
lack of model predictions with non-solar abundance ratios we decided to
avoid indices which are affected by overabundance problems (\eg\/ Mg \&
C$_2$4668). Instead we use here a combination of Fe-indices (Fe3) as
metal indicator which will bias our results towards the Fe abundance.
We note however, that our metallicities are not to be understood as
total metallicity but rather as a good estimate of the Fe-abundance.
Any non-solar abundance ratios which affect \HgA\/ are ignored. Model
predictions by W94 and V96 are overplotted in Figure~\ref{fig:fe3_hga}.

The ellipticals form a sequence of metallicity at roughly constant age.
The centres of the bright S0s NGC1380 \& NGC1381 follow the sequence of
Es. The remaining S0s cover a large range in metallicity and spread to
much younger luminosity weighted ages than the Es. We emphasize that
these age and metallicity estimates are {\em central luminosity
  weighted} estimates and for apparently young galaxies the derived
parameters are somewhat more insecure (see previous discussion about
the effects of composite stellar populations). The age and metallicity
gradients within the galaxies will be discussed in a future paper.
Fornax A, a bright peculiar S0, shows strong Balmer lines and strong Fe
absorption which translates into a luminosity weighted young and metal
rich stellar population. As we will see in the next section all other
young or metal poor S0s have velocity dispersions of $\sigma_0 \la
70$~\kms. The two galaxies with the weakest metal lines and strong
\Hb\/ \& \HgA\/ absorption (ESO359-G02, cross and ESO358-G25, open
triangle) appear to be different from the rest of the sample. These
galaxies are likely to be post-starburst or starburst galaxies
respectively. They have remarkable spectra for early-type galaxies,
showing blue continua, strong Balmer lines, and weak metal lines. These
galaxies are amongst the faintest in our sample and are $\sim$3\degr\/
away from the centre of the cluster (see also Kuntschner \& Davies
1998).

\section{LINE--STRENGTH INDICES AND THE CENTRAL VELOCITY DISPERSION}
\label{sec:linesigma}
The central velocity dispersion $\sigma_0$ of early-type galaxies is
known to correlate strongly with colours \cite{bow92} and the
absorption strength of the Mg-absorption feature at 5174~\AA\/
\cite{ter81,bur88,ben93,jor97,coll99}. The relatively small scatter
about these relations imply that the dynamical properties of galaxy
cores are closely connected with their stellar populations. However,
analysing the Mg--$\sigma_0$ relation for a sample of 736 mostly
early-type galaxies in 84 clusters, the EFAR group \cite{coll99} finds
a rather large dispersions in age (40\%) and in metallicity (50\%) at
fixed velocity dispersion using the constraints from the Mg--$\sigma_0$
relation {\em and} the Fundamental Plane. Correlations of other metal
indices, such as $<$Fe$>$, with the central velocity dispersion have
long been expected but so far relations have shown a large scatter and
only weak correlations \cite{fis96,jor97,jor99}. However, we will
demonstrate that galaxies in the Fornax cluster do show a clear
correlation between Fe-indices and central velocity dispersion.

Following Colless et~al. (1999) we find it more convenient to express
the ``atomic'' indices in magnitudes like the ``molecular'' index
Mg$_2$. The new index is denoted by the index name followed by a prime
sign [$^\prime$], \eg\/ \mgbp. Note that by using only the logarithm of
the atomic index, one introduces a non linear term in comparison to the
magnitude definition. Furthermore negative index values such as for the
\HgA\/ index cannot be put on a simple logarithmic scale. A priori it
is not clear whether $\log$~index or index$\,^\prime$ correlates better
with $\log \sigma_0$, but as the classical Mg--$\sigma_0$ relation was
established with Mg$_2$ measured in mag we adopt this approach here for
all other indices as well. The conversion between an index measured in
\AA\/ and magnitudes is

\begin{equation}
  \label{equ:mgbp}
  {\rmn index}\,^\prime= -2.5 \log \left( 1 - \frac{\rmn index}{\Delta \lambda}\right)  
\end{equation}
where $\Delta \lambda$ is the width of the index bandpass (see \eg\/
WO97 and Trager et~al. 1998 for a list of bandpass definitions).
Fe3$^\prime$ is defined as

\begin{equation}
  \label{equ:fe3p}
  {\rmn Fe3}^\prime = \frac{{\rmn Fe4383}^\prime + {\rmn Fe5270}^\prime +
  {\rmn Fe5335}^\prime}{3} \, .
\end{equation}

\begin{figure*}
\epsfig{file=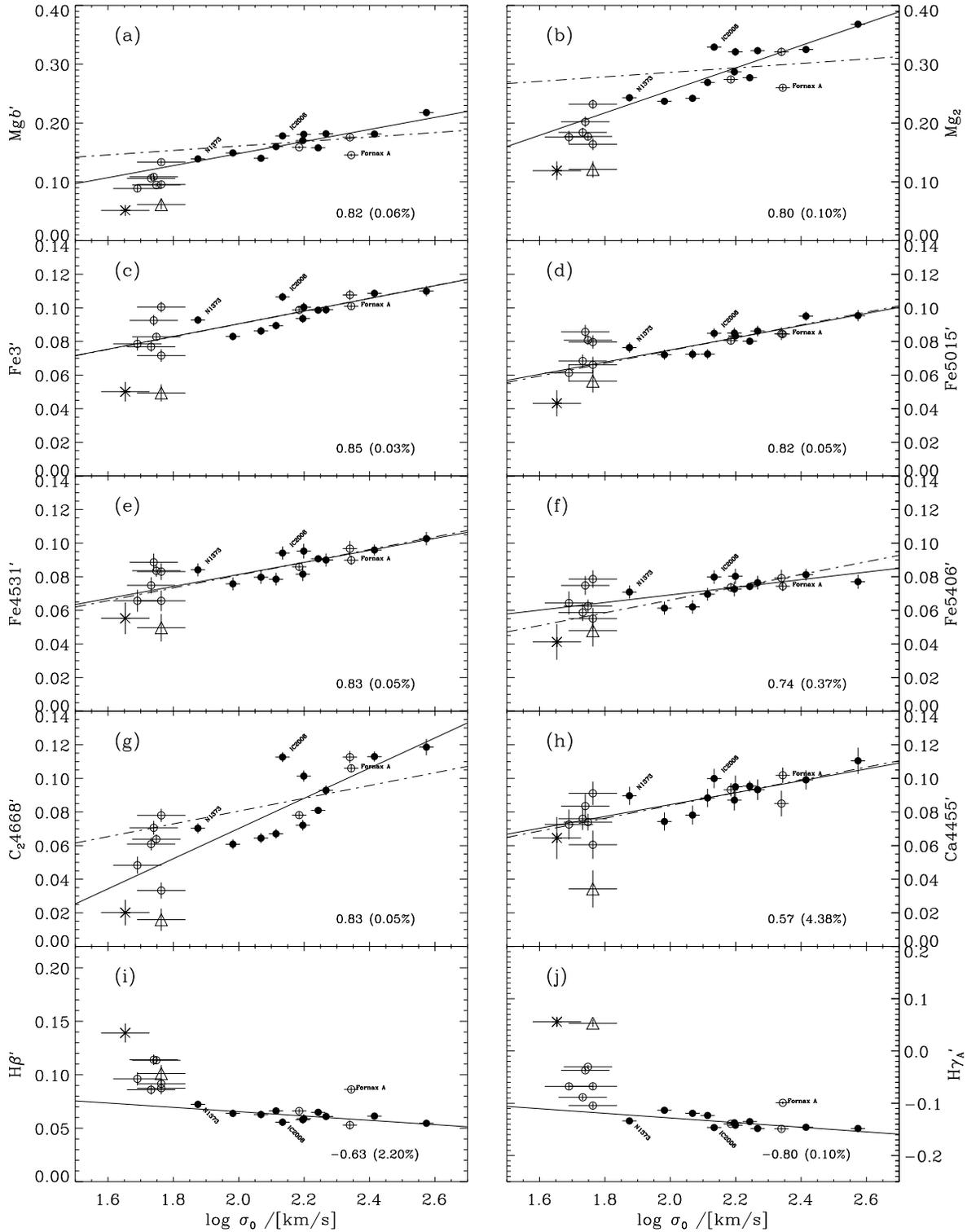,width=16.0cm}
\caption[]{\label{fig:allfe_sig}Selected metal absorption indices
  and Balmer line indices plotted against $\log \sigma_0$. All indices
  are measured in magnitudes following the conversion from
  Equation~\ref{equ:mgbp}. Note that all metal indices show a positive
  correlation and the Fe-indices show very similar slopes.  The
  dot-dashed line indicates a slope of 0.035 centred at $\log \sigma_0
  =2.2$. Note that the Mg indices exhibit a much steeper slope. The
  Spearman rank-order correlation coefficient for each dataset is shown
  with the significance level in brackets in the lower right corner of
  each panel. All Es and the large S0s NGC1380 \& NGC1381 were included
  in the fits.}
\end{figure*}

Figure~\ref{fig:allfe_sig} shows index--$\sigma_0$ relations for eight
different metal indices and two Balmer-line indices. The best fitting
linear relations and the scatter are summarized in
Table~\ref{tab:scaling} for all indices considered in this paper. For
the fits we used an ordinary least square method, minimizing the
residuals in y-direction \cite[hereafter OLS(Y$|$X)]{iso90}. Included
in the fit are all galaxies with old stellar populations, \ie\/ all Es
plus the bright S0s NGC1380 \& NGC1381; in total 13 galaxies. The
1-$\sigma$ scatter around the relation was robustly estimated by
deriving a value which includes 9 out of 13 galaxies (69\%). A
correlation coefficient derived from a (non parametric) Spearman
rank-order test is given in the lower right corner of each panel in
Figure~\ref{fig:allfe_sig}. The probability that the parameters are not
correlated is given in brackets.

%
%
%
\begin{table}
  \caption[]{Scaling relations}
  \label{tab:scaling} 
  \begin{tabular}{lcrl} \hline
index       &     &                                                     & scatter\\
            &     &                                                     & [mag]  \\
\hline
Mg$_2$      & $=$ & $( 0.191\pm0.023) \log \sigma_0 - ( 0.127\pm0.054)$ & 0.017\\
Mg$_1$      & $=$ & $( 0.136\pm0.015) \log \sigma_0 - ( 0.158\pm0.035)$ & 0.014\\
\mgbp       & $=$ & $( 0.102\pm0.020) \log \sigma_0 - ( 0.056\pm0.044)$ & 0.011\\
C$_2$4668   & $=$ & $( 0.090\pm0.018) \log \sigma_0 - ( 0.110\pm0.042)$ & 0.012\\
Fe3\prim    & $=$ & $( 0.038\pm0.011) \log \sigma_0 + ( 0.014\pm0.025)$ & 0.005\\
Fe4383\prim & $=$ & $( 0.043\pm0.019) \log \sigma_0 + ( 0.037\pm0.045)$ & 0.007\\
Fe4531\prim & $=$ & $( 0.036\pm0.010) \log \sigma_0 + ( 0.009\pm0.023)$ & 0.007\\
Fe5015\prim & $=$ & $( 0.036\pm0.008) \log \sigma_0 + ( 0.002\pm0.019)$ & 0.005\\
Fe5270\prim & $=$ & $( 0.029\pm0.009) \log \sigma_0 + ( 0.024\pm0.020)$ & 0.004\\
Fe5335\prim & $=$ & $( 0.043\pm0.009) \log \sigma_0 - ( 0.017\pm0.020)$ & 0.005\\
Fe5406\prim & $=$ & $( 0.023\pm0.012) \log \sigma_0 + ( 0.023\pm0.026)$ & 0.005\\
Ca4455\prim & $=$ & $( 0.035\pm0.017) \log \sigma_0 + ( 0.014\pm0.038)$ & 0.009\\
\Hb\prim    & $=$ & $-(0.020\pm0.007) \log \sigma_0 + ( 0.106\pm0.015)$ & 0.004\\
\HgA\prim   & $=$ & $-(0.045\pm0.019) \log \sigma_0 - ( 0.038\pm0.044)$ & 0.010\\
\HgF\prim   & $=$ & $-(0.049\pm0.017) \log \sigma_0 + ( 0.018\pm0.037)$ & 0.009\\ 
\hline
\multicolumn{4}{l}{Note -- Errors are estimated by a Jack-Knife error analysis.}
  \end{tabular} 
\end{table}

For the galaxies with old stellar populations the Mg--$\sigma_0$
relation is in excellent agreement with the literature (J{\o}rgensen
1997; Colless et~al. 1999). {\em Remarkably}, the Fe-line-indices also
show a clear positive correlation with the central velocity dispersion
and little scatter. This is the first time such strong correlations
have been found at a significant level. We note that all the Fe-line
and Ca4455--$\sigma_0$ relations show a slope consistent with a value
of $\sim0.035$. In contrast the slope of the Mg-lines and C$_2$4668 are
significantly steeper (see dot-dashed line in
Figure~\ref{fig:allfe_sig} and Table~\ref{tab:scaling}).

Although our Mg$_2$--$\sigma_0$ relation agrees well with the
literature values we find significant differences for other $\log
(index)$--$\sigma_0$ relations compared to the data of J{\o}rgensen
(1997,1999). Table~\ref{tab:scaling2} shows a comparsion of the slopes.
The $\log (<$Fe$>)$--$\sigma_0$ relation seems to be far steeper in the
Fornax cluster whereas the $\log \rmn{H}\beta_G$--$\sigma_0$ relation
is shallower compared to Coma. The $\log \rmn{C}_24668$--$\sigma_0$
relation in Fornax is marginally consistent with J{\o}rgensen (1997).
It is not clear why the $\log (index)$--$\sigma_0$ relations for
H$\beta_G$ and $<$Fe$>$ should be different to the Coma cluster. We
will present a possible explanantion at the end of this Section and in
Section~\ref{sec:discuss} where we discuss our results.

\begin{table}
  \caption[]{Scaling relations - comparsion of slopes with J{\o}rgensen (1997,1999)}
  \label{tab:scaling2} 
  \begin{tabular}{lccc} \hline
index                & this data         &  literature      & reference \\ 
\hline
$\log <$Fe$>$        &$( 0.209\pm0.047)$ &$( 0.075\pm0.025)$& (1)\\
                     &                   &$( 0.084\pm0.042)$& (2)\\
$\log \rmn{H}\beta_G$&$(-0.081\pm0.042)$ &$(-0.231\pm0.082)$& (1)\\
                     &                   &$(-0.169\pm0.038)$& (2)\\
$\log \rmn{C}_24668$ &$( 0.429\pm0.096)$ &$( 0.63 \pm0.06 )$& (1)\\
\hline
References:&\multicolumn{3}{l}{(1) J{\o}rgensen (1997, 11 nearby clusters)}\\
           &\multicolumn{3}{l}{(2) J{\o}rgensen (1999, Coma cluster)}\\
  \end{tabular} 
\end{table}

\begin{figure}
\epsfig{file=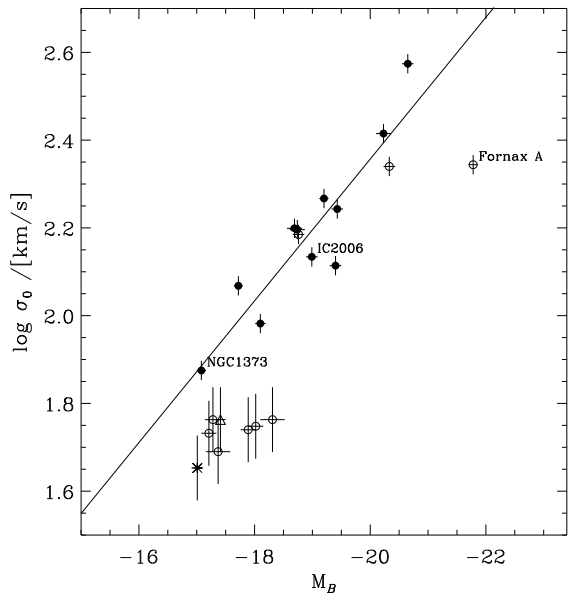, width=8.5cm}
\caption[]{\label{fig:MB_sig}The Faber--Jackson relation for the Fornax
  sample (assuming $m - M = 31.2$). The magnitudes are taken from the
  RC3 (de Vaucouleurs et~al. 1991). The best linear fit (OLS~(Y$|$X)
  including all Es, NGC1380 \& NGC1381) is shown as a solid line.}
\end{figure}

The centres of the two bright and old S0s NGC1380 \& NGC1381 follow
generally well the relation set by the elliptical galaxies. The lower
luminosity S0s have velocity dispersions $\sigma_0 \la 70$~\kms\/ and
show a large scatter about the mean relation of the old galaxies.
However, it is worth noting that they exhibit generally weak Mg
absorption and some of the faint S0s show as much Fe absorption as
L$^*$ ellipticals. Fornax~A, the brightest galaxy in our sample has a
central velocity dispersion of $\sigma_0 \simeq 220$~\kms\/ which is
too low compared to ellipticals of this luminosity in the Faber-Jackson
relation (see Figure~\ref{fig:MB_sig}). It also departs significantly
from the Mg--$\sigma_0$ relations in the sense that it shows too weak
Mg-absorption. As Fornax~A is regarded as the product of a recent
merger \cite{sch80,sch81,mac98} we interpret our results as strong
indications of at least one young stellar component in this galaxy.

One would expect the young stars in this galaxy to produce strong
Balmer absorption lines (as seen in Figure~\ref{fig:fe3_hga}) and to
dilute (or weaken) the metal lines of the underlying older stellar
component. However, if the burst mass is not too small, the relative
abundances of metal lines should in first order not be affected (see
discussion of composite stellar populations in
Section~\ref{sec:composite}). Yet, we find that Fornax~A deviates only
from the Mg--$\sigma_0$ relation and not from any of the other
metal-index--$\sigma_0$ relations (Figure~\ref{fig:allfe_sig}). We
interpret this as good evidence that the underlying older stellar
population of Fornax~A is significantly different from ellipticals at
this velocity dispersion, \ie the [Mg/Fe] ratio is lower, close to
solar.

Two of the ellipticals stand out from the normal metal
index--$\sigma_0$ relation: NGC1373 and IC2006 (labelled in
Figure~\ref{fig:allfe_sig}). These galaxies always show stronger metal
line absorption than what would be expected from the mean relation.
This is most prominent in the Fe3$^\prime$--$\sigma_0$ diagram
(panel~c). We note however, that the galaxies follow the mean
Faber-Jackson relation (Figure~\ref{fig:MB_sig}). There is little known
about the galaxy NGC1373; perhaps the best explanation why this
(elliptical) galaxy is somewhat off the mean relation is to regard it
as a transition galaxy between the sequence of Es and the faint S0s.
However, IC2006 has been studied in detail by Schweizer et~al. (1989).
They found a large counter-rotating ring of neutral hydrogen (HI)
associated with faint optical features and suggest that the HI ring may
have formed during a merger which created IC2006. Franx et~al. (1994)
re-analysed the optical photometry of Schweizer et~al. taking into
account the inclination of the galaxy and concluded that it probably
has a large disc in the outer parts which is seen almost face on and
therefore difficult to detect. They suggest that it should be
classified as E/S0 rather than a bona fide elliptical.

It seems plausible that the (perhaps peculiar) merger history of this
galaxy is the reason for its deviation from the index--$\sigma_0$
relations. However from our data, it is not clear whether the stellar
populations of IC2006 are too metal rich or whether the central
velocity dispersion is reduced compared to other elliptical galaxies of
this mass. If indeed this type of galaxy is more frequent in other
clusters, such as the Coma cluster (see J{\o}rgensen 1999), it would
explain why previous authors did not find a clear correlation of
Fe-lines with $\sigma_0$. A detailed analysis of the kinematics and
stellar population of this galaxy could be very valuable for our
understanding of how todays early-type galaxies were created.

Panels (i) \& (j) in Figure~\ref{fig:allfe_sig} show the
index--$\sigma_0$ relations for two Balmer lines. Both indices show
negative correlations. Elliptical galaxies and the bulges of NGC1380
and NGC1381 show little spread around the mean relation whereas the
younger galaxies, most remarkably NGC1316, tend to have significantly
stronger Balmer absorption at a given $\sigma_0$. We emphasize here
that the slope in the relation of the galaxies with old stellar
populations is mainly caused by a metallicity effect (metal poorer
galaxies have stronger Balmer absorption) and has little to do with age
differences. The two ``metal-rich'' galaxies IC2006 and NGC1373 are
deviant from the main \HgA$^\prime$--$\sigma_0$ relation in the sense
of lower \HgA\/ line strengths. This is caused by the residual
metallicity sensitivity of \HgA. The side-bands of this index are
located on metal lines which lower the pseudo-continuum level and thus
weaken the index.

\section{Global relations}
\label{sec:derived_sigma}
\begin{figure*}
  \epsfig{file=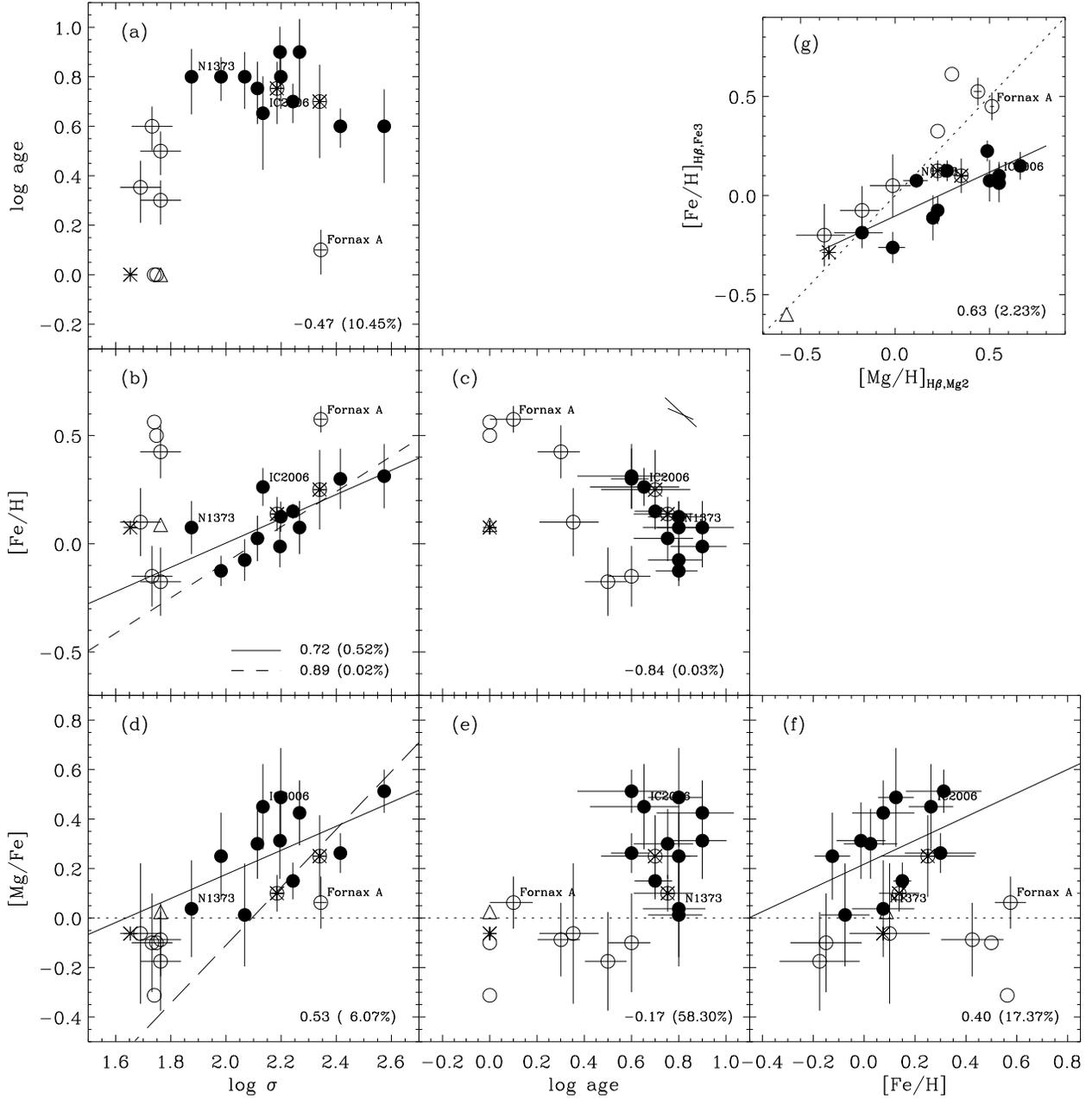,width=18cm}
\caption[]{\label{fig:age_metal_sig}The global relations between $\log
  {\rmn age}$, metallicity, [Mg/Fe] and $\log \sigma_0$ are shown. The
  filled circles represent elliptical galaxies and the open circles
  stand for the S0s. The two open circles with a cross indicate NGC1380
  and NGC1381. ESO358-G25 and ESO359-G02 are represented by a triangle
  and star respectively. For galaxies which lie at the edge or outside
  the model predictions in the age/metall icity diagnostic diagrams we
  do not show error bars. A Spearman rank-order correlation coefficient
  is shown in the lower right corner of each panel (significance in
  brackets). The Spearman rank-order test includes all galaxies with
  old stellar populations. The dashed line in panel (b) shows a linear
  fit excluding NGC1373 and IC2006. The long-dashed line in panel (d)
  shows the relation from J{\o}rgensen (1999) for the Coma cluster.
  Panel (g) shows the metallicity estimates derived from a \Hb\/ {\em
    vs}\/ Fe3 diagram (y-axis) plotted against the metallicity
  estimates derived from a \Hb\/ {\em vs}\/ Mg$_2$ diagram (x-axis).}
\end{figure*}

In this section we investigate the relations between our best age,
metallicity, [Mg/Fe] estimates and the central velocity dispersions.
Figure~\ref{fig:age_metal_sig} presents the results. The ages and
metallicities were estimated from a Fe3 {\em vs}\/ \HgA\/
age/metallicity diagnostic diagram (Figure~\ref{fig:fe3_hga}) in
combination with V96 models. The errors are evaluated following the
procedure outlined in Section~\ref{sec:non_solar} but only including
the Poisson error for individual galaxies. Some of the young galaxies
are at the edge or outside the range of the model predictions which
prevents an accurate error evaluation. For the latter galaxies we do
not plot error bars.  Notice that the ages and metallicities are
derived parameters which carry all the caveats discussed in the
previous sections. For example the independent measurement errors of
the line-strength indices translate into correlated errors in the age
-- metallicity plane due to the residual age/metallicity degeneracy in
the Fe3 {\em vs}\/ \HgA\/ diagram. We note that the results presented
in the following paragraphs would not change significanlty if \Hb\/ is
used as an age indicator (see Figure~\ref{fig:age_comp} and
\ref{fig:metal_comp} in the Appendix for a comparison of the age and
metallicity estimates derived from \HgA\/ \& \Hb). The Mg-overabundance
is estimated by evaluating the difference in metallicity estimate
between a Mg$_2$--\Hb\/ and a Fe3--\Hb\/ diagram (see
Figure~\ref{fig:mgfe_comp} in the Appendix for an estimation of [Mg/Fe]
using \HgA).

Our age estimates of ellipticals do not show a significant correlation
with $\log \sigma_0$ (panel~a). With the exception of Fornax~A all
galaxies with $\sigma_0 > 70$~\kms\/ show roughly the same age whereas
the younger galaxies populate the low velocity dispersion range.
However, there is a hint that the two dynamically hottest galaxies are
younger than their smaller brethren.

For galaxies with old stellar populations there is a clear correlation
between the central metallicity and the central velocity dispersion
$\sigma_0$ (panel~b). Consistent with our findings for the
Fe3--$\sigma_0$ relation the galaxies IC2006 and NGC1373 show a
stronger metal content than what would be expected from the mean
relation. The young S0s spread over the whole metallicity range. The
best fitting OLS(Y$|$X) relation (solid line, Jack-Knife error
analysis) to galaxies with old stellar populations gives:

\begin{equation}
  \label{eq:metal_sig2}
  {\rmn [Fe/H]} = (0.56\pm0.20) \log \sigma - (1.12\pm0.46) \, .
\end{equation}

A correlation coefficient derived from a Spearman rank-order test
(including all ellipticals and the two large S0s) is given in the lower
right corner of each panel in Figure~\ref{fig:age_metal_sig}. The
probability that the parameters are not correlated is given in
brackets. Excluding NGC1373 and IC2006 from the fit gives the following
relation (dashed line in panel b):

\begin{equation}
  \label{eq:metal_sig1}
  {\rmn [Fe/H]} = (0.82\pm0.18) \log \sigma - (1.72\pm0.40) \, .
\end{equation}

In the age-metallicity plane (panel~c) we find a statistically
significant relation in the sense that the more metal rich (and also
more luminous) galaxies are younger. The slope of this relation is
similar to what J{\o}rgensen (1999) found for the Coma cluster (see
also Worthey, Trager \& Faber 1995), yet the Fornax galaxies with
velocity dispersion $\sigma_0 > 70$~\kms\/ span a much smaller range in
age and [Fe/H]. We note that the non-treatment of non-solar abundance
ratios in combination with correlated errors could be the sole reason
for the trend found in Fornax. The direction and magnitude of
correlated errors for a galaxy of solar metallicity and 8~Gyrs age are
shown in panel (c), top right corner. Following on from the
age--metallicity relation J{\o}rgensen (1999) established for the Coma
cluster an age--[Mg/H]--$\sigma_0$ relation. It would be very
interesting to see whether such a correlation exists also in Fornax.
However, the small number of galaxies combined with a rather small
spread in age makes such an analysis very insecure and has therefore
not been attempted.

The Mg-overabundance shows a weak positive correlation with central
velocity dispersion and [Fe/H] in the sense that dynamically hotter and
more metal rich galaxies are more overabundant (panel d, f). In the
Fornax cluster significant overabundances are found for galaxies with
$\sigma_0 \ga 100$~\kms\/ or $\rmn{[Fe/H]} \ga 0.0$ (panel~f, g). The
best fitting linear relation between [Mg/Fe] and $\log \sigma_0$ is:

\begin{equation}
  \label{eq:mgfe_sig}
  {\rmn [Mg/Fe]} = (0.49\pm0.18) \log \sigma - (0.80\pm0.41) \, .
\end{equation}

This relation is qualitatively in agreement with the results from the
Coma cluster (J{\o}rgensen 1999, long-dashed line in
Figure~\ref{fig:age_metal_sig}d). The scatter about the
[Mg/Fe]--$\sigma_0$ relation in Fornax is consistent with the errors
for [Mg/Fe], but there seems to be a rather large spread in the [Mg/Fe]
ratio at a given metallicity (panel f).  Although the latter is in good
agreement with our findings in the Fe-Mg$_2$ diagram
(Figure~\ref{fig:mg_fe}), we note that the errors are heavily
correlated in the [Mg/Fe]--[Fe/H] diagram. There is no significant
correlation of the [Mg/Fe] ratio with $\log {\rmn age}$ where the young
S0s show solar or slightly less than solar [Mg/Fe] ratios.

\section{DISCUSSION}
\label{sec:discuss}
In this study, great care was taken to calibrate the line-strength
measurements to a standard system in which we can compare the results
with theoretical model predictions (Section~\ref{sec:lick_cal}). The
accuracy of this calibration is vital when one wants to derive absolute
age and metallicity estimates. Although we were able to demonstrate the
high quality of our calibration some unresolved issues such as the
systematic offset in the Mg$_2$ {\em vs}\/ \mgb\/ diagram, the rather
large rms error in the original Lick/IDS stellar library and perhaps
most important of all the largely unknown effects of non-solar
abundance ratios prevent us from deriving accurate absolute age and
metallicity estimates. However, for the discussion of {\em relative}
differences in the stellar populations of early-type galaxies our data
set and current models are of excellent use.

In this paper we have made use of two stellar population models
provided by Worthey (1994) and Vazdekis et~al. (1996). Both models make
use of the Lick/IDS fitting functions but have otherwise somewhat
different prescriptions to predict line-strength indices of integrated
single-burst stellar populations (SSP). The predictions of the two
models are consistent and our conclusions would not change if only one
of them had been used for the analysis. To our knowledge, this would be
also true if we had used any other model which makes use of the
Lick/IDS fitting functions.

One of the most important results from this study is the homogeneity of
the stellar populations in dynamically hot early-type galaxies in the
Fornax cluster. Apart from Fornax~A all early-type galaxies (Es \& S0s)
with $\sigma_0 > 70$~\kms\/ are of roughly the same age and their
central metallicity scales with $\log \sigma_0$. The homogeneity is
reflected in tight relations of observables such as Mg--$\sigma_0$ and
Fe--$\sigma_0$ and a clear correlation of [Fe/H] with the central
velocity dispersion. The existence of the latter is reassuring in terms
of our current understanding of the colour-magnitude-relation (CMR) in
clusters being mainly a result of increasing metallicity with
increasing luminosity \cite{kod97,ter99}.

Previous authors \cite{fis96,jor97,jor99} pointed out that the lack of
a correlation of Fe-absorption strength with central velocity
dispersion would give evidence for a second parameter or conspiracy of
age, metallicity and [Mg/Fe] ratio which keeps the CMR tight. For
example in the Coma cluster J{\o}rgensen (1999) did not find a strong
correlation of $<$Fe$>$ with central velocity dispersion and hence her
[Fe/H]--$\sigma_0$ relation is also not significant. However, both the
$<$Mg$>$--$\sigma_0$ and [Mg/H]--$\sigma_0$ relation are clearly seen
in Coma. In this context it is important to note that the slope of the
Fe--$\sigma_0$ relation which one would expect from the change of
metallicity in the CMR \cite{kod97} is quite shallow and therefore only
detectable with high S/N data. In contrast, the Mg--$\sigma_0$ relation
is steeper, and therefore easier to detect. The reason for this is a
combination of a larger dynamical range in the Mg indices compared to
the average Fe-index and an increasing Mg overabundance with central
velocity dispersion giving a steeper slope than what would be expected
from the change in metallicity only.

An alternative explanation for the lack of a Fe--$\sigma_0$ relation in
Coma could be based on galaxies such as IC2006, which do not follow the
Fe--$\sigma_0$ relation very well. If this type of galaxy is more
frequent in the Coma cluster than in Fornax it would be impossible to
find a clear Fe--$\sigma_0$ relation. In summary we find that in the
Fornax cluster there is no need for a second parameter such as age,
metallicity or [Mg/Fe] to keep the CMR tight. Indeed, we favour an
interpretation where small variations of age, metallicity and/or
[Mg/Fe] at any given $\sigma_0$ are responsible for some real scatter
in the scaling relations for the Fornax cluster. However, we emphasize
that this may not be true for other (larger?) clusters.

In addition to the population of old, dynamically hot early-type
galaxies, we find a sizeable fraction of young, dynamically colder
($\sigma_0 \la 70$~\kms) systems within our magnitude limited survey.
Some of the young S0s (NGC1375, ESO359-G02 and ESO358-G25) fit in
remarkably well with the predictions of galaxy harassment in clusters
(Moore, Lake, Katz 1998; Lake, Katz, Moore 1998). In this scenario,
medium sized disc galaxies (Sc-type) fall into a cluster environment
and get ``harassed'' by high speed encounters with cluster galaxies.
The end-products are small spheroidal galaxies where some gas of the
disc is driven into the centre of the galaxy. This gas is likely to be
turned into stars in a central stellar burst. We note, that most of
these young galaxies are in the periphery of the Fornax cluster
consistent with having been ``accreted'' onto the cluster from the
field.

Two of the S0s which show young populations in the centre, also have
extended discs (NGC1380A and IC1963). This seems to be in contradiction
with the harassment picture. However we emphasize, that the existing
harassment simulations do not include spirals with a substantial bulge
component. Here the bulge is likely to stabilize the disc and the
end-products may be able to keep substantial disc components (Ben
Moore, private communication). The existence of a population of
dynamically colder galaxies with young stellar populations in the
nuclear regions is in agreement with a typical (nearby) cluster CMR
where one finds a tail of blue galaxies towards the faint end (\eg\/
Terlevich 1998). Furthermore Terlevich et~al. (1999) demonstrate for
the Coma cluster, using line-strength analysis, that these blue
galaxies contain young stellar populations rather than being metal
poor.

It seems that in the Fornax cluster significant amounts of young
stellar populations are predominantly found in low luminosity
(lenticular) systems. However, for a sample of Coma cluster early-type
galaxies Mehlert (1998) found that relatively bright S0s spread over
the whole range in age (Es, excluding the cDs, are found to be old).
This of course raises the question whether morphology is the driving
parameter for young stellar populations (only S0s are younger) or
whether luminosity is the important parameter (low luminosity E \& S0
galaxies are on average younger). Taking the results from Coma and
Fornax together we would like to argue that in clusters it is only the
lenticular galaxies which show signs of recent star formation and that
low luminosity lenticular systems are more likely to do so. The latter
may be just caused by the recent accretion of these low luminosity
systems onto the cluster.

So far we have addressed the age and metallicity distributions in the
Fornax cluster with the help of line-strength indices. However there is
more detailed information on the star formation (SF) processes to be
gained if one investigates the [Mg/Fe] abundance ratios. When new stars
are formed chemical enrichment is predominantly driven by the ejecta of
SN~Ia (main producer of Fe peak elements) and SN~II (producing mainly
alpha elements). However, SN~Ia are delayed compared to SN~II which
explode on short time-scales of $\le10^6-10^7$ yr. Taking this into
account there are mainly two mechanisms which determine the Mg/Fe ratio
in galaxies: (i) the star formation time scale and (ii) the fraction of
high mass stars, \ie\/ the initial mass function (IMF) (see \eg\/
Worthey et~al. 1992). As re-confirmed in this study, the majority of
cluster early-type galaxies show a trend of increasing [Mg/Fe] ratio
with central velocity dispersion. Galaxies with young stellar
populations and/or low luminosity galaxies show roughly solar abundance
ratios. Given that the most luminous galaxies are also the metal
richest, we emphasize that any realistic star-formation models have to
be able to produce metal rich and Mg-overabundant stars at the same
time.

In principle one can reproduce the observed trends of overabundances
and metallicity with varying star-formation time-scales: {\em large}
galaxies form within {\em shorter} time-scales than smaller galaxies
\cite{bre96}. However, this leeds to extremely short star-formation
time-scales for the most massive galaxies. A plausible way to solve
this dilemma would be a varying IMF where massive galaxies have a top
heavy IMF and low luminosity galaxies show a more Salpeter like IMF.
For further discussions of the matter see also Peletier (1999) and
Tantalo, Chiosi \& Bressan (1998).

Recently, Thomas \& Kauffmann (1999, see also Thomas 1999) presented
preliminary results from their semi-analytic galaxy formation models
for the distribution of [Mg/Fe] in galaxies as a function of
luminosity. In this scenario luminous ellipticals are the last to form
and hence Thomas \& Kauffmann find a trend that the [Mg/Fe] ratio
decreases with increasing luminosity, opposite to the observed trend.
In general it seems very difficult with the current stellar population
models to reproduce the observed Magnesium strength, and therefore
[Mg/Fe] values ($\rmn{[Mg/Fe]} \sim 0.4$) of luminous ellipticals
(Greggio 1997, but also see Sansom \& Proctor 1998).

\section{CONCLUSIONS}
\label{sec:conclusion}
We have measured the {\em central} line strength indices in a magnitude
limited sample of early-type galaxies brighter than ${\rmn M}_B=-17$ in
the Fornax cluster and have applied the models of Worthey (1994),
Worthey \& Ottaviani (1997) and Vazdekis~et~al. (1996) to estimate
their ages, metallicities and abundance ratios. We find that:

\begin{enumerate}
\item {\bf Elliptical Galaxies} appear to be roughly coeval forming a
  sequence in metallicity varying roughly from $-0.25$ to $0.30$ in
  [Fe/H].  This result is consistent with the conventional view of old,
  coeval elliptical galaxies where the metallicity scales with the
  luminosity of the galaxy. This is reflected in scaling relations such
  as Mg--$\sigma_0$. Remarkably, we could show that all other metal
  line-strength indices also clearly correlate with the central
  velocity dispersion. In fact all Fe-line--$\sigma_0$ relations are
  consistent with having the same slope.
  
\item {\bf Lenticular Galaxies} have luminosity weighted metallicities
  spanning the whole range of SSP model predictions. Lower luminosity
  S0s show {\em luminosity weighted ages} younger than those of the
  ellipticals. However, the centres of the bright lenticular galaxies
  NGC1380 and NGC1381 resemble the properties of ellipticals suggesting
  that they experienced similar star formation histories. The peculiar
  S0 galaxy Fornax A (NGC1316), which is the brightest galaxy in the
  sample, has strong Balmer lines implying a very young luminosity
  weighted age, yet the metallicity is equal to the most metal rich Es.
  This is consistent with Fornax~A having been involved in a recent
  gaseous merger. The S0s NGC1380 and NGC1381 follow the
  index--$\sigma_0$ relations of the ellipticals very well. However,
  the S0s with a young stellar component generally show a large scatter
  around the scaling relations.
  
\item Our conclusions are based on several age/metallicity diagnostic
  diagrams which give consistent results. Furthermore we demonstrate
  the advantage of using an emission robust age indicator such as
  H$\gamma_{\rm{A}}$ when analysing the stellar populations of
  extragalactic objects.
  
\item We have discovered that two of the fainter and very metal poor
  lenticular galaxies appear to have undergone major star-formation in
  the last 2 Gyrs (in one case very much more recently). We note that,
  like Fornax A, most of the young galaxies lie on the periphery of the
  cluster. This is consistent with the harassment picture where these
  galaxies are accreted from the field and undergo a morphological
  transformation with a central star burst.
  
\item The elliptical galaxies and the S0 NGC1380 exhibit overabundances
  up to 0.4 dex in Magnesium compared to Fe. There is a trend that the
  most massive and metal rich galaxies are the most overabundant,
  whereas the fainter Es approach solar ratios. This trend is
  inconsistent with the currently available semi-analytical predictions
  for hierarchical galaxy formation. S0s with young stellar populations
  are consistent with roughly solar abundance ratios, and may even be
  slightly underabundant. Remarkably also Fornax~A, the brightest
  galaxy in our sample, shows close to solar abundance ratios which is
  not what one would expect of an early-type galaxy of its size.
  
\item Furthermore we note that abundance ratio trends, which are not
  included in the models, can lead to a change of relative age and
  metallicity estimates depending on which index combination is used in
  the analysis. As long as the non-solar abundance ratios are not
  properly incorporated into the models the estimation of absolute ages
  of integrated stellar populations remains insecure.

\end{enumerate}

\section*{ACKNOWLEDGEMENTS}
HK acknowledges the use of STARLINK computing facilities at the
University of Durham and also wishes to thank the
Dr.~Carl~Duisberg~Stiftung and University of Durham for generous
financial support during the course of this work. Special thanks go to
Roger Davies who provided excellent supervision throughout this
project. Interesting and very helpful discussions of the following
people are acknowledged: Eric Bell, Richard Bower, Taddy Kodama,
Reynier Peletier, Jim Rose and Alexandre Vazdekis. Thanks also go to
Scott Trager and Guy Worthey for providing the Lick/IDS measurements of
the higher order Balmer lines. HK thanks the referee, Inger
J{\o}rgensen, for a careful and thorough reading of this paper which
helped to improve the final presentation.

\appendix
\section{Kinematics}
\label{app:kin_cal}
\begin{figure}
    \epsfig{file=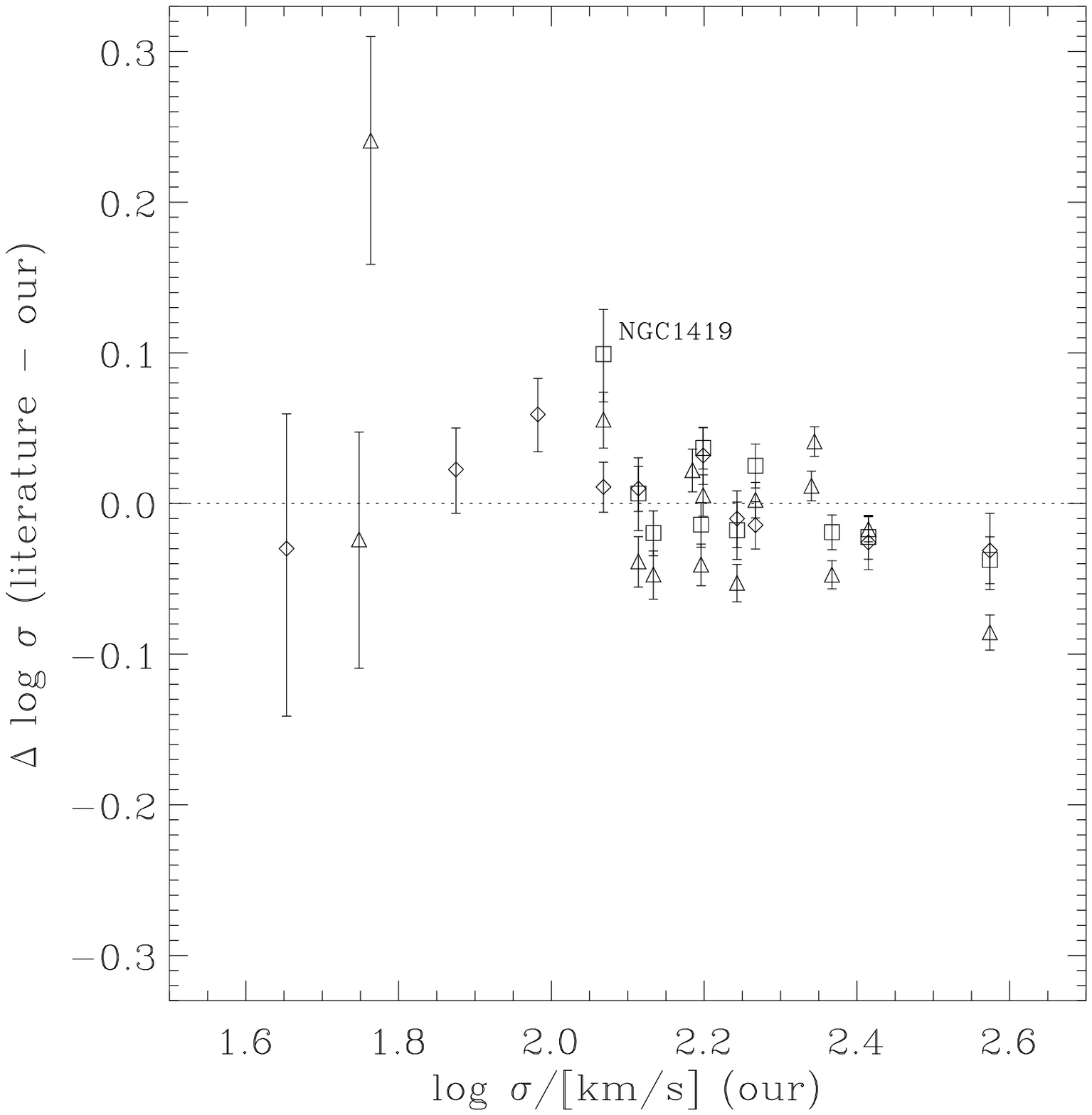,width=8.5cm}
\caption[]{\label{fig:comp_sig}Comparison of our central velocity
  dispersions with McElroy (1995, triangles), Smith (1998, squares) and
  Graham et~al. (1998, diamonds). Error bars reflect the literature
  errors and the rms scatter due to different template stars for the
  AAT data.}
\end{figure}

In Figure~\ref{fig:comp_sig} a literature comparison of the central
velocity dispersion measurements is presented. The most recent data for
Fornax galaxies from Graham et~al. (1998) as well as the literature
compilation of Smith (1998, Private communication) is in good agreement
with our data. The comparison with the literature compilation by
McElroy (1995) shows a somewhat larger scatter. In order to establish
an average error for the velocity dispersion estimates the mean scatter
of the data compared to Smith (1998) and Graham et~al. (1998) is
evaluated. For the comparison all galaxies with $\sigma_0 < 70$~\kms\/
and also NGC1419 (marked in Figure~\ref{fig:comp_sig}) from the Smith
compilation were excluded. The mean scatter is 0.027 in $\log
\sigma_0$. Subtracting in quadrature a mean error of 0.015 for the
literature data gives an error of 0.022 [in $\log \sigma_0$] for
galaxies with $\sigma_0 \ge 70$~\kms. For galaxies with $\sigma_0
<70$~\kms\/ we adopt the rms scatter of the template stars as estimate
of the error ($\Delta \log \sigma_0 = 0.074$).

\section{Lick/IDS calibration} 
\label{app:lick_cal}
\begin{figure*}
\epsfig{file=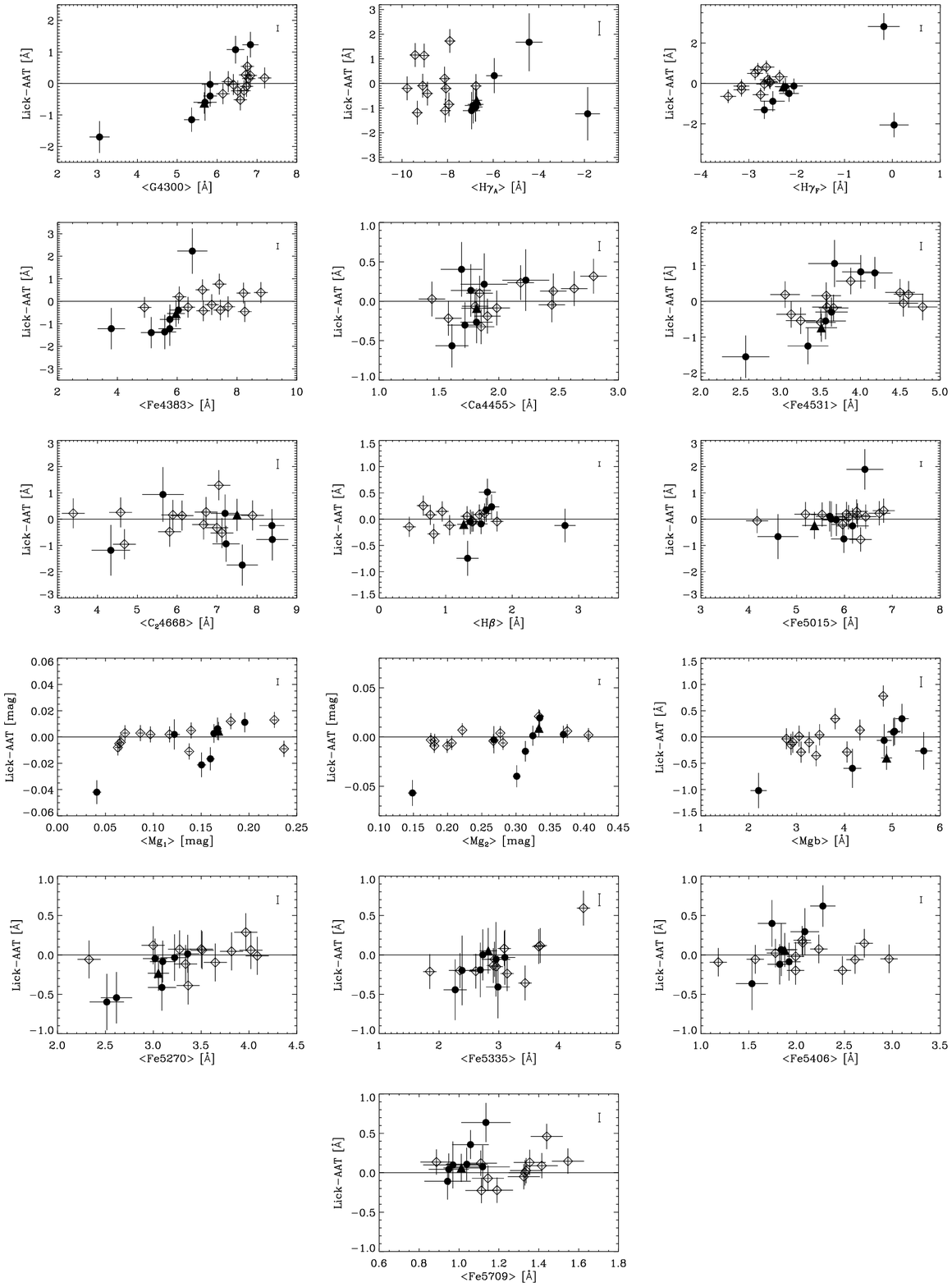,width=15.5cm}
\caption[]{\label{fig:lick_off}Comparison of 16 indices between
  Lick/IDS and our measurements {\em after} the Lick/IDS offset
  correction (see Table~\ref{tab:lick_off}) has been applied. Open
  diamonds, filled circles and the filled triangle represent stars,
  Fornax-galaxies and NGC3379 respectively. The formal error in the
  offset is shown as an error bar in the upper right corner of each
  panel.}
\end{figure*}

In Figure~\ref{fig:lick_off} a comparison between the original Lick/IDS
index measurements of stars and selected galaxies in common with our
data is shown. The individual diagrams show the scatter
(Lick/IDS~-~AAT) around the mean offset {\em vs}\/ the average of
Lick/IDS and AAT.

\begin{figure*}
\epsfig{file=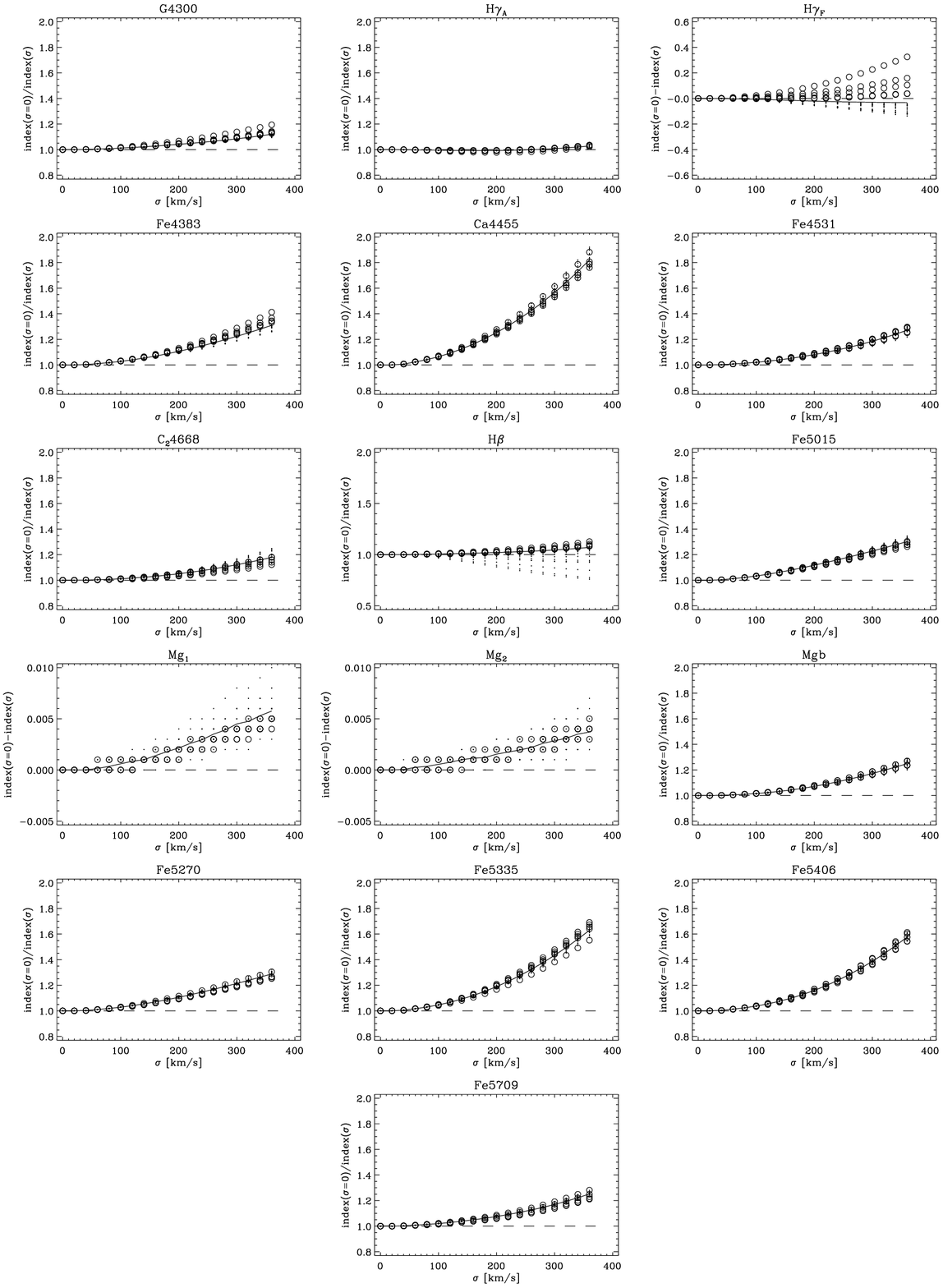,width=15.5cm}
\caption[]{\label{fig:vel_disp_corr}Velocity dispersion
  corrections. Dots and open circles represent stars and galaxies
  respectively. The solid line connects the mean of all data points in
  each $\sigma$-bin. Note that for \Hb\/ only stars with equivalent
  width of \Hb\/ $>1.1$ \AA\/ have been used to determine the mean. See
  text for details.}
\end{figure*}

Figure~\ref{fig:vel_disp_corr} shows the velocity dispersion
corrections for each index as derived from broadened stellar and
selected galaxy spectra.

\section{Comparison of derived ages, metallicities and [Mg/Fe] ratios}
\label{sec:comp_stellar}
\begin{figure}
  \epsfig{file=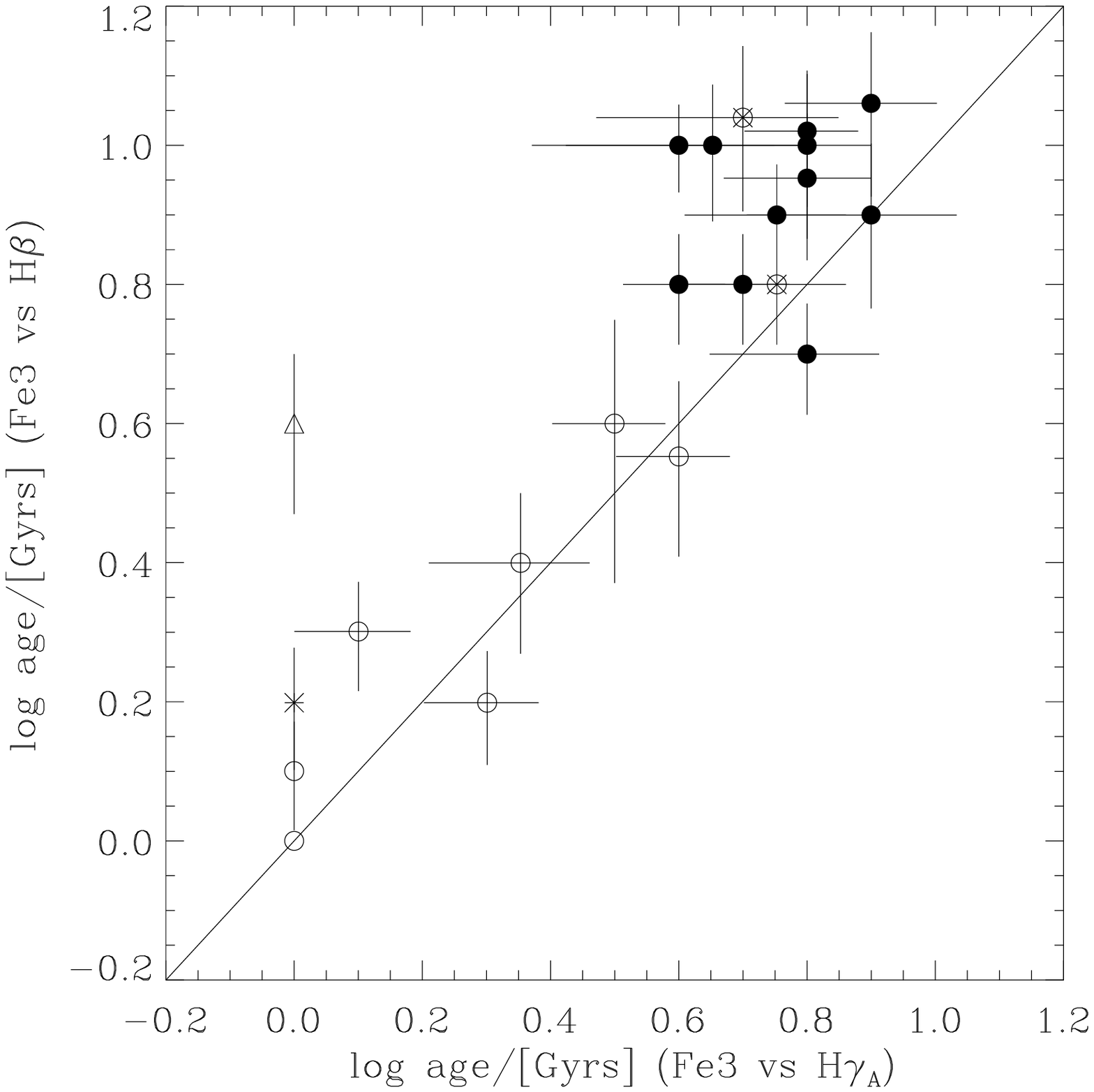,width=8.5cm}
\caption[]{\label{fig:age_comp}Comparison of the ages derived from a Fe3
  {\em vs} \HgA\/ diagram and a Fe3 {\em vs} \Hb\/ diagram using the
  models of Vazdekis (1996). For details of the method see
  Section~\ref{sec:non_solar}.}
\end{figure}

\begin{figure}
    \epsfig{file=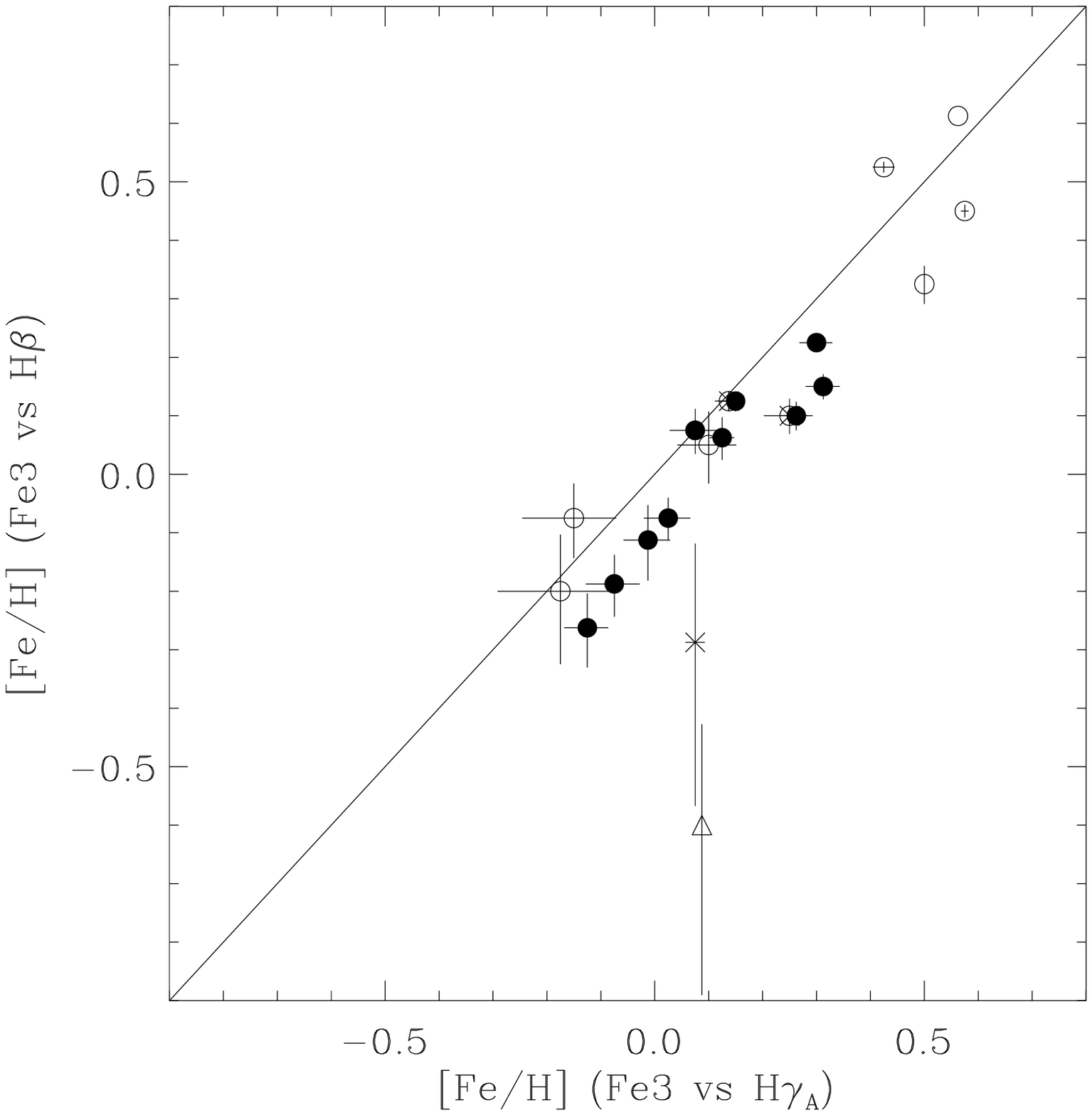,width=8.5cm}
\caption[]{\label{fig:metal_comp}Comparison of the metallicities derived from a Fe3
  {\em vs} \HgA\/ diagram and a Fe3 {\em vs} \Hb\/ diagram using the
  models of Vazdekis (1996). For details of the method see
  Section~\ref{sec:non_solar}.}
\end{figure}

\begin{figure}
    \epsfig{file=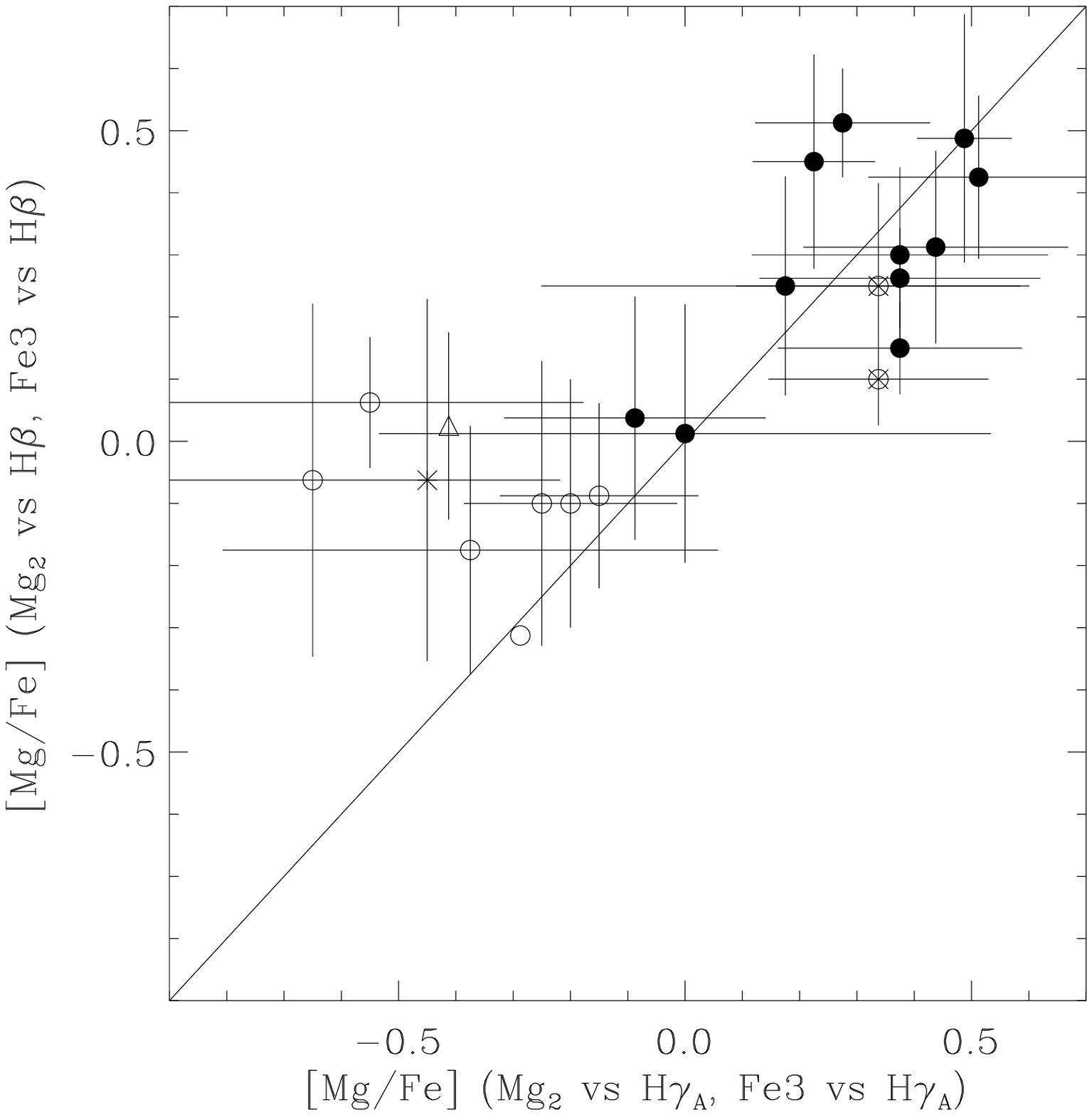,width=8.5cm}
\caption[]{\label{fig:mgfe_comp}Comparison of the [Mg/Fe] ratios derived
  from a Fe3 {\em vs} \HgA, Mg$_2$ {\em vs} \HgA\/ diagram and a Fe3
  {\em vs} \Hb, Mg$_2$ {\em vs} \Hb\/ diagram using the models of
  Vazdekis (1996). For details of the method see
  Section~\ref{sec:non_solar}.}
\end{figure}

Here we present three Figures (\ref{fig:age_comp}, \ref{fig:metal_comp}
\& \ref{fig:mgfe_comp}) comparing our age, metallicity and [Mg/Fe]
estimates derived with different index combinations. In general we find
very consistent results when we use \eg\/ \Hb\/ instead of \HgA\/ as an
age indicator. Most of the differences, such as the [Mg/Fe] ratios of
some of the young lenticular galaxies, is caused by the data points
lying at the edge of the model predictions or where they have been
extrapolated. The extrapolation was necessary because the models
reflect roughly solar abundance ratios and cannot account for \eg\/ the
strongest Mg$_2$ absorption found in our sample.

\section{Final calibrated central indices}
\subsection{Table of central index measurements}
The final, fully corrected central (2\farcs$\times$3\farcs85) index
measurements and associated errors for the Fornax galaxies and NGC3379
are presented in Table~D2. For each galaxy we give our Lick/IDS index
measurement in the first row and the 1$\sigma$ error in the second row.
The second column (first part of Table~D2) lists central velocity
dispersions in $\log \sigma$ units. The last column (part two of
Table~D2) lists H$\beta_{\rmn G}$ (not Lick/IDS index); for further
details see Section~\ref{sec:hb_gon}.

\subsection{\Hb\/ {\em vs}\/ H$\beta_{\rmn G}$ relation}
\label{sec:hb_gon}
\begin{figure}
\epsfig{file=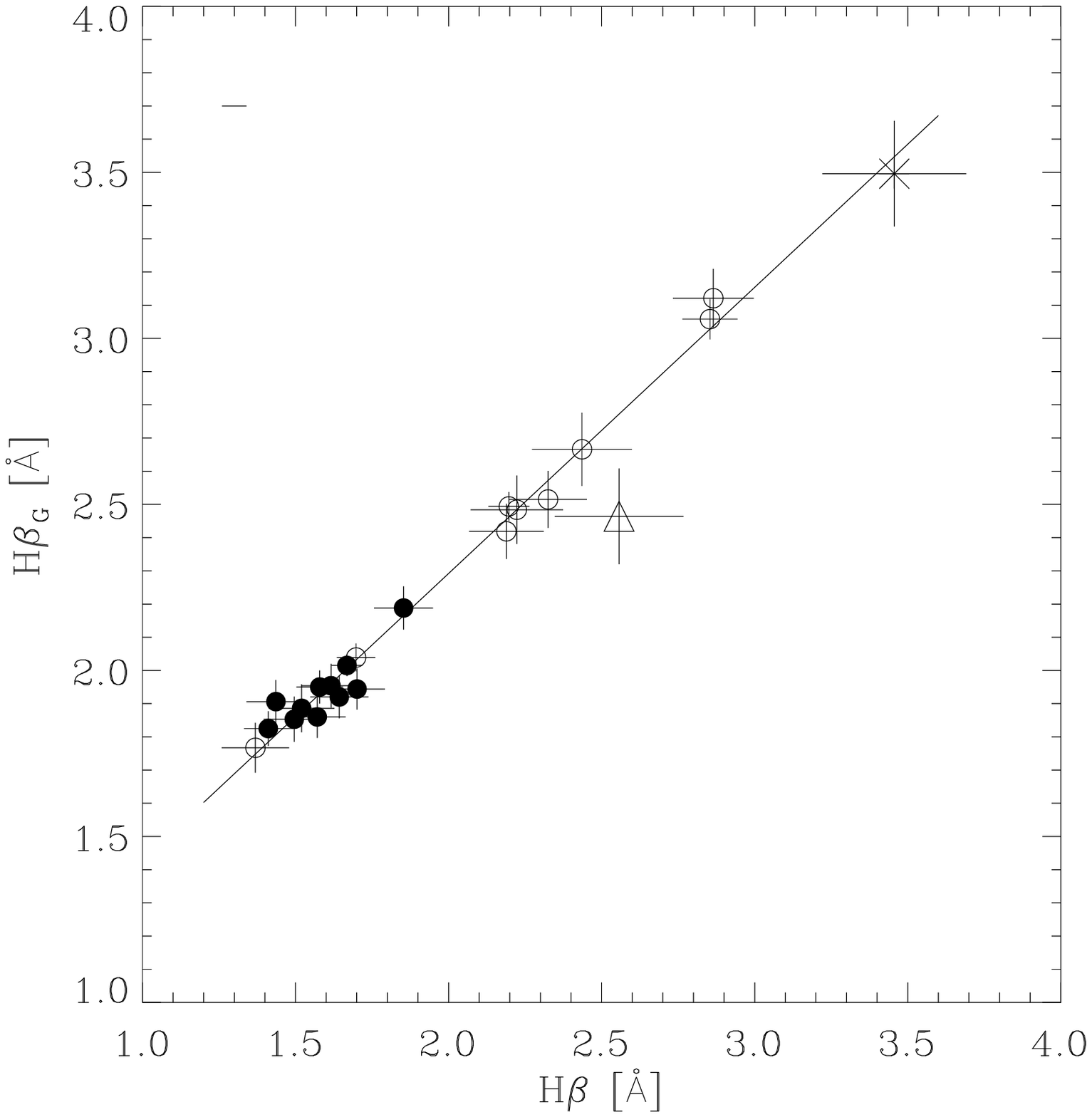, width=8.5cm}
\caption[]{\label{fig:hb_hbg}The relation between the Lick/IDS \Hb\/ index and
  H$\beta_{\rmn G}$ as defined in J{\o}rgensen (1997). The fit includes
  all galaxies but ESO358-G25 (open triangle) because of its emission
  contamination. The error bar in the upper left corner shows the rms
  error in the offset of the \Hb\/ index to the Lick/IDS system.}
\end{figure}

Figure~\ref{fig:hb_hbg} shows a plot of \Hb\/ equivalent width {\em
  vs}\/ H$\beta_{\rmn G}$ equivalent width. The bandpasses of the
H$\beta_{\rmn G}$ index are defined by J{\o}rgensen (1997), based on an
earlier definition of an \Hb\/ emission index by Gonz\'{a}lez (1993), in
such a way that the influence of the Fe feature right next to the \Hb\/
absorption feature is minimized. For wavelength definitions see
Table~\ref{tab:hbg}.

\begin{table}
\caption[]{Wavelength definition of H$\beta_{\rmn G}$}
\label{tab:hbg}
\begin{tabular}[h]{lll}
\hline
Index               & Index bandpass     &  Pseudocontinua  \\ \hline
H$\beta_{\rmn G}$   & 4851.32 -- 4871.32 &  4815.00 4845.00 \\
                    &                    &  4880.00 4930.00 \\ \hline
\end{tabular}
\end{table}

The indices \Hb\/ \& H$\beta_{\rmn G}$ show an excellent correlation
for the Fornax sample. The solid line in Figure~\ref{fig:hb_hbg} shows
the fit to all the data, excluding ESO358-G25 (open triangle) because
of its emission contamination, using a method which bisects the
ordinary least squares fits made by minimising the X and the Y
residuals:

\begin{equation}
  \label{eq:hb_hbg}
  {\rmn H}\beta_{\rmn G} = (0.862\pm0.027) \cdot {\rmn H}\beta + (0.568\pm0.051)
\end{equation}
The derived relation (errors from a Jack-Knife analysis) is in very
good agreement with the relation found by J{\o}rgensen (1997).

\begin{table*}
\begin{minipage}{80mm}
\caption[]{Fully corrected Lick/IDS indices for the central (2\farcs$\times$3\farcs85) extraction} 
\label{tab:indices_gal}
\begin{tabular}{lrrrrrrrrr}  \hline 
  Name    &log $\sigma$&G4300&Fe4383&Ca4455&Fe4531&C4668&\Hb  &Fe5015&Mg$_1$\\
          &            &[\AA]&[\AA] &[\AA] &[\AA] &[\AA]&[\AA]&[\AA] &[mag]   \\\hline
  NGC1316 &  2.344     &5.14 & 5.95 & 2.02 & 3.57 & 8.02&2.20 & 5.70 & 0.119\\
  ~~~$\pm$&  0.022     &0.11 & 0.16 & 0.09 & 0.12 & 0.19&0.07 & 0.16 & 0.003\\
  NGC1336 &  1.982     &5.47 & 4.85 & 1.49 & 3.03 & 4.70&1.64 & 4.90 & 0.097\\
  ~~~$\pm$&  0.022     &0.15 & 0.21 & 0.11 & 0.16 & 0.23&0.09 & 0.21 & 0.005\\
  NGC1339 &  2.199     &5.93 & 6.15 & 1.88 & 3.78 & 7.69&1.52 & 5.63 & 0.168\\
  ~~~$\pm$&  0.022     &0.17 & 0.25 & 0.14 & 0.18 & 0.26&0.11 & 0.25 & 0.006\\
  NGC1351 &  2.196     &6.04 & 5.56 & 1.73 & 3.26 & 5.54&1.50 & 5.74 & 0.137\\
  ~~~$\pm$&  0.022     &0.16 & 0.23 & 0.13 & 0.17 & 0.25&0.10 & 0.23 & 0.005\\
  NGC1373 &  1.875     &5.98 & 5.83 & 1.78 & 3.36 & 5.41&1.85 & 5.18 & 0.101\\
  ~~~$\pm$&  0.022     &0.15 & 0.21 & 0.11 & 0.16 & 0.24&0.10 & 0.21 & 0.005\\
  NGC1374 &  2.267     &6.21 & 5.80 & 1.85 & 3.58 & 7.07&1.57 & 5.82 & 0.164\\
  ~~~$\pm$&  0.022     &0.15 & 0.22 & 0.13 & 0.16 & 0.24&0.09 & 0.22 & 0.005\\
  NGC1375 &  1.748     &3.90 & 4.40 & 1.48 & 3.33 & 4.93&2.85 & 5.47 & 0.062\\
  ~~~$\pm$&  0.074     &0.14 & 0.19 & 0.10 & 0.15 & 0.22&0.09 & 0.20 & 0.005\\
  NGC1379 &  2.114     &5.99 & 5.37 & 1.76 & 3.14 & 5.16&1.70 & 4.93 & 0.121\\
  ~~~$\pm$&  0.022     &0.14 & 0.21 & 0.11 & 0.16 & 0.23&0.09 & 0.21 & 0.005\\
  NGC1380 &  2.340     &5.92 & 6.26 & 1.70 & 3.84 & 8.50&1.37 & 5.72 & 0.161\\
  ~~~$\pm$&  0.022     &0.18 & 0.27 & 0.16 & 0.19 & 0.29&0.11 & 0.27 & 0.006\\
 NGC1380A &  1.740     &4.16 & 5.04 & 1.67 & 3.53 & 5.43&2.87 & 5.79 & 0.076\\
  ~~~$\pm$&  0.074     &0.20 & 0.28 & 0.15 & 0.21 & 0.32&0.13 & 0.29 & 0.007\\
  NGC1381 &  2.185     &6.09 & 5.82 & 1.85 & 3.42 & 5.99&1.70 & 5.45 & 0.121\\
  ~~~$\pm$&  0.022     &0.11 & 0.15 & 0.08 & 0.11 & 0.16&0.06 & 0.15 & 0.003\\
  NGC1399 &  2.574     &5.90 & 6.49 & 2.18 & 4.06 & 8.93&1.41 & 6.41 & 0.191\\
  ~~~$\pm$&  0.022     &0.17 & 0.29 & 0.16 & 0.17 & 0.39&0.08 & 0.24 & 0.004\\
  NGC1404 &  2.415     &6.03 & 6.26 & 1.96 & 3.81 & 8.52&1.58 & 6.39 & 0.162\\
  ~~~$\pm$&  0.022     &0.14 & 0.20 & 0.12 & 0.14 & 0.24&0.08 & 0.20 & 0.004\\
  NGC1419 &  2.068     &5.75 & 5.28 & 1.56 & 3.18 & 4.97&1.62 & 4.92 & 0.111\\
  ~~~$\pm$&  0.022     &0.15 & 0.22 & 0.12 & 0.16 & 0.24&0.10 & 0.22 & 0.005\\
  NGC1427 &  2.243     &5.95 & 5.92 & 1.89 & 3.61 & 6.20&1.67 & 5.43 & 0.127\\
  ~~~$\pm$&  0.022     &0.09 & 0.12 & 0.07 & 0.09 & 0.14&0.05 & 0.12 & 0.003\\
   IC1963 &  1.763     &5.08 & 6.03 & 1.81 & 3.31 & 5.98&2.33 & 5.39 & 0.093\\
  ~~~$\pm$&  0.074     &0.19 & 0.27 & 0.14 & 0.21 & 0.31&0.13 & 0.28 & 0.007\\
   IC2006 &  2.134     &5.95 & 6.18 & 1.98 & 3.74 & 8.50&1.44 & 5.73 & 0.171\\
  ~~~$\pm$&  0.022     &0.15 & 0.22 & 0.12 & 0.16 & 0.23&0.10 & 0.22 & 0.005\\
 E359-G02 &  1.653     &1.64 & 2.77 & 1.30 & 2.23 & 1.58&3.46 & 2.97 & 0.039\\
  ~~~$\pm$&  0.074     &0.35 & 0.50 & 0.26 & 0.39 & 0.61&0.23 & 0.55 & 0.013\\
 E358-G06 &  1.763     &5.00 & 4.05 & 1.22 & 2.64 & 2.60&2.22 & 4.51 & 0.057\\
  ~~~$\pm$&  0.074     &0.23 & 0.33 & 0.17 & 0.25 & 0.38&0.15 & 0.34 & 0.008\\
 E358-G25 &  1.763     &1.58 & 2.28 & 0.70 & 2.01 & 1.25&2.56 & 3.86 & 0.039\\
  ~~~$\pm$&  0.074     &0.30 & 0.43 & 0.23 & 0.34 & 0.53&0.21 & 0.47 & 0.012\\
 E358-G50 &  1.690     &4.85 & 4.22 & 1.46 & 2.64 & 3.76&2.44 & 4.18 & 0.064\\
  ~~~$\pm$&  0.074     &0.24 & 0.35 & 0.18 & 0.27 & 0.41&0.16 & 0.36 & 0.009\\
 E358-G59 &  1.732     &5.04 & 4.84 & 1.52 & 3.00 & 4.71&2.19 & 4.66 & 0.067\\
  ~~~$\pm$&  0.074     &0.19 & 0.26 & 0.14 & 0.20 & 0.30&0.12 & 0.27 & 0.007\\\hline
  NGC3379 &  2.367     &6.02 & 6.30 & 1.90 & 3.92 & 7.48&1.33 & 5.56 & 0.166\\
  ~~~$\pm$&  0.022     &0.15 & 0.23 & 0.13 & 0.16 & 0.24&0.09 & 0.21 & 0.004\\ \hline
\end{tabular}
\end{minipage}
\end{table*}
\begin{table*}
\begin{minipage}{80mm}
\contcaption{}
\begin{tabular}{lrrrrrrrrr}  \hline 
  Name    &Mg$_2$&\mgb &Fe5270&Fe5335&Fe5406&Fe5709&\HgA &\HgF &$\mbox{H}\beta_G$\\  
          &[mag] &[\AA]&[\AA] &[\AA] &[\AA] &[\AA] &[\AA]&[\AA]& [\AA]       \\\hline
  NGC1316 &0.260 & 4.08& 3.10 & 2.90 & 1.82 & 1.04 &-4.18&-0.64&   2.49      \\      
  ~~~$\pm$&0.004 & 0.08& 0.08 & 0.10 & 0.07 & 0.05 & 0.11& 0.08&   0.04      \\      
  NGC1336 &0.237 & 4.17& 2.70 & 2.33 & 1.51 & 0.81 &-4.82&-1.34&   1.92      \\      
  ~~~$\pm$&0.006 & 0.10& 0.12 & 0.13 & 0.10 & 0.08 & 0.17& 0.10&   0.06      \\      
  NGC1339 &0.321 & 4.99& 3.04 & 2.72 & 1.96 & 0.93 &-6.11&-1.91&   1.89      \\      
  ~~~$\pm$&0.007 & 0.12& 0.13 & 0.16 & 0.11 & 0.09 & 0.20& 0.12&   0.07      \\      
  NGC1351 &0.287 & 4.72& 3.03 & 2.51 & 1.78 & 0.98 &-5.89&-1.85&   1.85      \\      
  ~~~$\pm$&0.006 & 0.11& 0.12 & 0.15 & 0.11 & 0.09 & 0.18& 0.11&   0.07      \\      
  NGC1373 &0.243 & 3.90& 2.95 & 2.29 & 1.74 & 0.89 &-5.74&-1.55&   2.19      \\      
  ~~~$\pm$&0.006 & 0.10& 0.12 & 0.13 & 0.10 & 0.08 & 0.17& 0.11&   0.06      \\      
  NGC1374 &0.323 & 5.01& 3.13 & 2.76 & 1.87 & 0.98 &-6.40&-2.02&   1.86      \\      
  ~~~$\pm$&0.006 & 0.10& 0.12 & 0.14 & 0.10 & 0.08 & 0.17& 0.11&   0.06      \\      
  NGC1375 &0.177 & 2.71& 2.88 & 2.49 & 1.54 & 1.08 &-1.24& 1.06&   3.06      \\      
  ~~~$\pm$&0.006 & 0.10& 0.11 & 0.13 & 0.10 & 0.08 & 0.14& 0.09&   0.06      \\      
  NGC1379 &0.269 & 4.45& 2.80 & 2.47 & 1.71 & 0.92 &-5.26&-1.59&   1.94      \\      
  ~~~$\pm$&0.006 & 0.10& 0.11 & 0.13 & 0.10 & 0.08 & 0.17& 0.10&   0.06      \\      
  NGC1380 &0.321 & 4.86& 3.24 & 3.18 & 1.93 & 1.00 &-6.43&-2.16&   1.77      \\      
  ~~~$\pm$&0.007 & 0.13& 0.14 & 0.17 & 0.13 & 0.09 & 0.20& 0.13&   0.08      \\      
 NGC1380A &0.202 & 3.09& 3.01 & 2.84 & 1.83 & 1.01 &-1.52& 0.96&   3.12      \\      
  ~~~$\pm$&0.009 & 0.15& 0.16 & 0.19 & 0.14 & 0.12 & 0.21& 0.13&   0.09      \\      
  NGC1381 &0.274 & 4.42& 3.15 & 2.71 & 1.80 & 0.97 &-5.99&-1.87&   2.04      \\      
  ~~~$\pm$&0.005 & 0.07& 0.08 & 0.09 & 0.07 & 0.05 & 0.12& 0.08&   0.04      \\      
  NGC1399 &0.368 & 5.91& 3.36 & 3.11 & 1.88 & 0.83 &-6.40&-2.04&   1.83      \\      
  ~~~$\pm$&0.004 & 0.16& 0.11 & 0.17 & 0.11 & 0.07 & 0.12& 0.14&   0.05      \\      
  NGC1404 &0.325 & 5.00& 3.37 & 3.14 & 1.98 & 0.88 &-6.29&-1.99&   1.95      \\      
  ~~~$\pm$&0.004 & 0.10& 0.10 & 0.13 & 0.09 & 0.06 & 0.13& 0.10&   0.05      \\      
  NGC1419 &0.242 & 3.93& 2.72 & 2.30 & 1.53 & 0.82 &-5.08&-1.50&   1.95      \\      
  ~~~$\pm$&0.006 & 0.10& 0.12 & 0.14 & 0.10 & 0.08 & 0.17& 0.11&   0.07      \\      
  NGC1427 &0.277 & 4.40& 3.13 & 2.64 & 1.82 & 0.99 &-5.80&-1.68&   2.02      \\      
  ~~~$\pm$&0.003 & 0.06& 0.06 & 0.07 & 0.05 & 0.04 & 0.09& 0.07&   0.03      \\      
   IC1963 &0.232 & 3.76& 3.06 & 2.82 & 1.92 & 1.04 &-4.41&-0.88&   2.52      \\      
  ~~~$\pm$&0.008 & 0.14& 0.15 & 0.17 & 0.13 & 0.11 & 0.22& 0.14&   0.09      \\      
   IC2006 &0.329 & 4.92& 3.19 & 3.17 & 1.95 & 0.96 &-6.32&-2.13&   1.91      \\      
  ~~~$\pm$&0.006 & 0.10& 0.12 & 0.14 & 0.10 & 0.08 & 0.18& 0.11&   0.06      \\      
 E359-G02 &0.119 & 1.50& 1.87 & 1.38 & 1.02 & 0.77 & 2.19& 2.69&   3.50      \\      
  ~~~$\pm$&0.016 & 0.28& 0.31 & 0.36 & 0.27 & 0.22 & 0.33& 0.20&   0.16      \\      
 E358-G06 &0.164 & 2.74& 2.54 & 1.95 & 1.36 & 0.60 &-2.81&-0.07&   2.48      \\      
  ~~~$\pm$&0.010 & 0.17& 0.19 & 0.22 & 0.16 & 0.14 & 0.25& 0.15&   0.10      \\      
 E358-G25 &0.121 & 1.78& 1.99 & 1.57 & 1.19 & 0.52 & 2.09& 1.92&   2.46      \\      
  ~~~$\pm$&0.014 & 0.24& 0.28 & 0.32 & 0.24 & 0.20 & 0.28& 0.17&   0.14      \\      
 E358-G50 &0.176 & 2.55& 2.74 & 2.34 & 1.58 & 0.95 &-2.82& 0.00&   2.67      \\      
  ~~~$\pm$&0.011 & 0.18& 0.20 & 0.23 & 0.17 & 0.15 & 0.27& 0.16&   0.11      \\      
 E358-G59 &0.184 & 3.02& 2.30 & 2.11 & 1.45 & 0.89 &-3.72&-0.49&   2.42      \\      
  ~~~$\pm$&0.008 & 0.13& 0.15 & 0.17 & 0.13 & 0.11 & 0.21& 0.13&   0.08      \\\hline
  NGC3379 &0.329 & 5.13& 3.21 & 2.86 & 1.89 & 0.99 &-6.39&-2.20&   1.80      \\      
  ~~~$\pm$&0.005 & 0.11& 0.11 & 0.14 & 0.10 & 0.07 & 0.16& 0.11&   0.06      \\\hline

\end{tabular}
\end{minipage}
\end{table*}

\bsp

\label{lastpage}

\end{document}